%% file: pandora.tex
\newif\ifNotes\Notestrue

\documentclass[sigconf,nonacm,screen]{acmart}

\usepackage[normalem]{ulem}

\usepackage[hyphenbreaks]{breakurl}

\input{builtins/std}

\definecolor{Green}{HTML}{00A64F}
\definecolor{NavyBlue}{HTML}{006EB8}

\definecolor{darkviolet}{HTML}{9400D3}
\definecolor{darkgreen}{HTML}{006400}

\ifNotes
\newcommand{\colorcomment}[2]{\leavevmode\unskip\space{\color{#1}#2}\xspace}
\newcommand{\taggedcolorcomment}[3]{\colorcomment{#1}{[\textbf{#2}: #3]}}
\else
\newcommand{\colorcomment}[2]{\leavevmode\unskip\relax}
\newcommand{\taggedcolorcomment}[3]{\leavevmode\unskip\relax}
\fi

\definecolor{tocitecolor}{rgb}{0.65,0,0}

\definecolor{nightmode}{HTML}{E7E5E4}
\definecolor{nightmodetext}{HTML}{17191C}

\newcommand{\tool}{\textsc{Lm\-Test}\xspace}
\newcommand{\dsl}{\textsc{Lm\-Spec}\xspace}
\newcommand{\hook}[1]{\mathop{\mathbf{#1}}}

\usepackage{paralist}
\usepackage{multirow}
\usepackage{float}
\usepackage{booktabs}
\usepackage{algorithm}
\usepackage[noend]{algpseudocode}
\usepackage{nth}
\usepackage{fontawesome}
\usepackage{enumitem}

\usepackage{tikz}
\usetikzlibrary{shapes.geometric, arrows}

\usepackage[frozencache=true,cachedir=_minted-pandora]{minted} 
\newminted[hyblock]{hylang}{linenos=none, breaklines=true, fontsize=\small}
\newcommand{\hycode}[1]{\texttt{#1}}

\usepackage{scalerel}

\usepackage{stackengine}
\def\rlwd{.4pt}
\def\rlht{3pt}
\def\shatvrule{\rule{\rlwd}{\rlht}}
\def\shat#1{%
  \setbox0=\hbox{\!\,\!#1\!\,\!}%
  \stackon[0pt]{\stackon[-1.5pt]{{#1}}{%
    \shatvrule\kern\wd0\kern-\rlwd\kern-\rlwd\shatvrule}}%
    {\rule{\wd0}{\rlwd}}%
}

\newcommand{\conf}{s}

\newcommand{\tup}[1]{\langle #1 \rangle}

\newcommand{\compactparagraph}[1]{\smallskip\noindent\textbf{#1:\ }}

\newcommand{\LeakageModel}{\ensuremath{\mathit{LM}}\xspace}

\newcommand{\SilentStores}{\mathbf{SS}}
\newcommand{\RegisterCompression}{\mathbf{RFC}}

\newcommand{\ValueRegisterCompressionZero}{\mathbf{RFC0}}
\newcommand{\NarrowRegisterCompression}{\mathbf{RFCN}}
\newcommand{\ComputationReuse}{\mathbf{CR}}

\newcommand{\TrivialComputationSimplification}{\mathbf{CST}}

\newcommand{\ComputationReuseAddress}{\mathbf{CRA}}
\newcommand{\CacheCompression}{\mathbf{CC}}

\newcommand{\Seq}{S\textsc{eq}\xspace}
\newcommand{\PHT}{P\textsc{ht}\xspace}
\newcommand{\SLS}{S\textsc{ls}\xspace}
\newcommand{\STL}{S\textsc{tl}\xspace}
\newcommand{\RSB}{R\textsc{sb}\xspace}
\newcommand{\RSBdrop}{R\textsc{sb}$_\bot$\xspace}
\newcommand{\RSBcirc}{R\textsc{sb}$_\circ$\xspace}

\newcommand{\found}[1]{$\bullet$}

\newcommand{\instr}[1]{#1\xspace}

\crefname{section}{\S}{\S\S}
\Crefname{section}{\S}{\S\S}
\crefformat{section}{\S#2#1#3}
\creflabelformat{section}{#2\textup{#1}#3}

\begin{document}

\title{Testing side-channel security of cryptographic implementations against future microarchitectures}

\author{Gilles Barthe}
\affiliation{%
  \institution{Max Planck Institute  \\ for Security and Privacy (MPI-SP)}
  \city{Bochum}
  \country{Germany}
}
\additionalaffiliation{IMDEA Software Institute}
\email{gilles.barthe@mpi-sp.org}

\author{Marcel B\"ohme}
\affiliation{%
  \institution{Max Planck Institute  \\for Security and Privacy (MPI-SP)}
  \city{Bochum}
  \country{Germany}
}
\email{marcel.boehme@mpi-sp.org}

\author{Sunjay Cauligi}
\affiliation{%
  \institution{Max Planck Institute \\for Security and Privacy (MPI-SP)}
  \city{Bochum}
  \country{Germany}
}
\email{sunjay.cauligi@mpi-sp.org}

\author{Chitchanok Chuengsatiansup}
\affiliation{%
  \institution{The University of Melbourne}
  \city{Melbourne}
  \country{Australia}
}
\email{c.chuengsatiansup@unimelb.edu.au}

\author{Daniel Genkin}
\affiliation{%
  \institution{GeorgiaTech}
  \city{Atlanta}
  \country{United States}
}
\email{genkin@gatech.edu}

\author{Marco Guarnieri}
\affiliation{%
  \institution{IMDEA Software Institute}
  \city{Madrid}
  \country{Spain}
}
\email{marco.guarnieri@imdea.org}

\author{David Mateos Romero}
\affiliation{%
  \institution{IMDEA Software Institute}
  \city{Madrid}
  \country{Spain}
}
\email{david.mateos.romero@gmail.com}

\author{Peter Schwabe}
\affiliation{%
\institution{Max Planck Institute \\for Security and Privacy (MPI-SP)}
\city{Bochum}
\country{Germany}
}
\additionalaffiliation{Radboud University}
\email{peter@cryptojedi.org}

\author{David Wu}
\affiliation{%
  \institution{University of Adelaide}
  \city{Adelaide}
  \country{Australia}
}
\email{david.wu@adelaide.edu.au}

\author{Yuval Yarom}
\affiliation{%
  \institution{Ruhr University Bochum}
  \city{Bochum}
  \country{Germany}
}
\email{yuval.yarom@rub.de}

\renewcommand{\shortauthors}{G. Barthe, M. B\"ohme, S. Cauligi, C. Chuengsatiansup, D. Genkin, M. Guarnieri, D. Mateos Romero, P. Schwabe, D. Wu, Y. Yarom}

\input{abstract}

\maketitle

\input{intro}

\input{overview}

\input{formal}

\input{testing}

\input{eval}

\input{defenses}
\input{discussion}

\input{related}

\input{conclusion}

\bibliographystyle{ACM-Reference-Format}
\bibliography{bib}

\begin{appendix}
  \input{leakagemodels}

\end{appendix}

\end{document}

%% file: builtins/std.tex
\usepackage{amsmath,amsfonts,amsthm} %
\usepackage{comment}                         %
\usepackage{ifthen}                          %
\usepackage{minibox}                         %
\usepackage{multirow}                        %
\usepackage{suffix}                          %
\usepackage{graphicx}                        %
\usepackage{url}                    %
\usepackage{xargs}                           %
\usepackage{xcolor}              %
\usepackage{hyperref} %
\usepackage[capitalise,noabbrev,nameinlink]{cleveref}            %
\usepackage{hyphenat}                        %
\usepackage{fancyvrb}                        %
\usepackage{supertabular}                    %
\usepackage{xspace}                          %
\usepackage{microtype}                       %

\hypersetup{breaklinks=true}

\newcommand{\ignore}[1]{}

\def\dash---{\kern.16667em---\penalty\exhyphenpenalty\hskip.16667em\relax}

  {\begin{list}{$\blacktriangleright$}%
    {\leftmargin=\parindent \itemsep=2pt \topsep=2pt
     \parsep=0pt \partopsep=0pt}}%
  {\end{list}}

%% file: abstract.tex
\begin{abstract}

How will future microarchitectures impact the security of existing cryptographic implementations?
As we cannot keep  
reducing the size of transistors,
chip vendors have started developing new microarchitectural optimizations to speed up computation. 
A recent study (Sanchez Vicarte et al., ISCA 2021) suggests that these optimizations might open the Pandora's box of microarchitectural attacks. 
However, there is little guidance on how to evaluate the security impact of future optimization proposals.

To help chip vendors explore the impact of microarchitectural optimizations on cryptographic implementations, we develop (i)~an expressive domain-specific language, called \textsc{LmSpec}, that allows them to specify the leakage model for the given optimization and (ii)~a testing framework, called \textsc{LmTest}, to automatically detect leaks under the specified leakage model within the given implementation.
Using this framework, we conduct an empirical study of 18 proposed microarchitectural optimizations on 25 implementations of eight cryptographic primitives in five popular  libraries. We find that \emph{every} implementation would contain secret-dependent leaks, sometimes sufficient to recover a victim's secret key, if these optimizations were realized.
Ironically, some leaks are possible only \emph{because} of coding idioms used to \emph{prevent} leaks under the standard constant-time model.\looseness=-1

    \end{abstract}

%% file: intro.tex
\section{Introduction}

As reducing the size of transistors is increasingly more challenging due to physical limits, 
chip vendors are looking into microarchitectural optimizations as an alternative to further speed up computations.
These optimizations exploit the spatial and temporal locality exhibited by programs to predict future behavior and perform some of the computation in advance.
Thus, the processor can cache or prefetch data in anticipation of a near-future use, or it can predict future instruction flow and even computation results in an effort to better exploit the processor's inherent parallelism.

However, a recent study~\cite{pandorasbox} suggests that this surge of new
micro\-archi\-tectures opens the Pandora's box of \emph{microarchitectural
attacks}~\cite{GeYCH18}, which
exploit the side effects of microarchitectural optimizations on a
program's execution time to compromise the confidentiality of
an otherwise secure computation. 
Although the impact of existing
 optimizations, like
caching~\cite{primeprobe,flushreload} or speculative execution
\cite{spectre}, is well understood and captured by secure programming
guidelines~\cite{almeida2016verifying,pitchfork}, there is little
guidance on how to evaluate the security impact of possible future
microarchitectures.

As a first step in this direction, \citet{pandorasbox} 
performed a systematic review of
recent optimizations proposed by the computer architecture community
and suggested that many of those would likely leak secret information
in unexpected ways. However, the authors left open how exactly these
optimizations would impact existing cryptographic implementations and whether existing mitigations, like leakage-resistent
coding idioms~\cite{slni, BernsteinLS12}, would help in preventing leaks. %

To substantiate this prognosis, in this paper we develop a framework for assessing the side-channel guarantees of programs against  (future) microarchitectural optimizations.
Our framework consists of (1)~\dsl{}, an expressive domain-specific language that allows chip vendors and software developers to specify the leakage model associated with  a given optimization in terms of leakage traces, and (2)~\tool{}, a testing approach that generates random inputs and automatically detects secret-dependent leaks for the specified \dsl{} leakage models within a given cryptographic implementation.
We  use this framework to (3) perform a large scale study of the impact of the optimization proposals studied in~\cite{pandorasbox} on the side-channel guarantees of mainstream cryptographic implementations.
Next, we overview these contributions in more detail.

\compactparagraph{\dsl{} language}
We develop the \dsl{} language for specifying leakage models at the ISA-level.
Following the formalism of Guarnieri et al.~\cite{contracts}, we define an \dsl{} leakage model as (a)~a \emph{leakage clause} (\Cref{sect:leakage-models}), which specifies what observations are leaked during the execution of a program, and (b)~a \emph{prediction clause} (\Cref{sect:execution-models}), which specifies the prediction mechanisms supported by the microarchitecture and what their effects are.
Hence, \dsl{} models map each program execution to a \emph{leakage trace} capturing the leaked information.
Importantly, models specified in \dsl{} are \emph{executable} for the x86 ISA (\Cref{sect:testing-approach:dsl-executable-semantics}).
That is, given an initial program state, we can execute an arbitrary x86 binary and derive the leakage trace generated by the \dsl{} model.
To achieve this, \dsl{} models are automatically translated into handlers for the Unicorn CPU emulator~\cite{unicorn} capturing the relevant information to generate leakage traces for a program execution.

\compactparagraph{\tool{} testing tool}
We develop the \tool{} testing framework for detecting leaks in x86 programs with respect to a given \dsl{} leakage model (\Cref{sect:testing-approach}).
\tool{} takes as input the binary of the program under test, an entry point in the program together with labels for each program input (indicating whether the parameter is public or secret), and the \dsl{} model.
To detect leaks, \tool{} adopts a relational testing approach~\cite{revizor}:
(1) it randomly generates test cases consisting of \emph{pairs} of initial program states that are \emph{low-equivalent}, i.e., that differ only in the value of secret inputs, (2) it then executes the program on both inputs in a test case to derive the leakage traces according to the \dsl{} model, and (3) it finally checks for differences in the two leakage traces in a test case.
A test case demonstrating different leakage traces indicates a leak of secret information.
We remark that differently from other random relational testing approaches~\cite{he2020ct,revizor}, which target \textit{fixed} leakage models, \tool{} is parametric in the \dsl{} leakage model, thereby allowing one to study the security implications of any optimization proposal whose leakage profile can be represented in \dsl{}.

\compactparagraph{Case study}
To evaluate our framework and to assess the risks that proposed optimizations pose to current software, we report on a
large-scale empirical study of the impact of microarchitectural
optimizations on popular cryptographic libraries (\Cref{sect:case-studies}). We consider 18
optimizations, under six different execution models, and 25
cryptographic implementations selected from five popular cryptographic libraries,
including libsodium and rust-crypto. 
We find that \emph{every}
implementation would contain secret-dependent leaks if these
optimizations were realized.
In some cases, an op\-ti\-mi\-za\-tion-induced leak would be sufficient to recover a victim's secret key directly from the leakage trace (e.g., for the X25519 implementation in libsodium~\cite{libsodium}). Ironically, some leaks are possible \emph{only because} of coding idioms, such as constant-time swap or bit-masking, used to \emph{prevent} leaks under the standard constant-time model.

\compactparagraph{Summary of contributions}
In summary, the paper makes the following contributions:
(1)~the \dsl{} language for rapidly prototyping leakage models (\S\ref{sect:leakage-models}--\ref{sect:execution-models}),
(2)~the \tool{} testing tool for detecting secret-dependent leaks in programs against an arbitrary \dsl{} model (\Cref{sect:testing-approach}), and 
(3)~a large-scale case study analyzing the side-channel guarantees of mainstream cryptographic implementations against recent microarchitectural proposals (\Cref{sect:case-studies}).

\compactparagraph{Artifacts}
The implementation of \tool{} and \dsl{}, together with all leakage and prediction clauses from our case study as well as scripts to reproduce all our results are available at \url{https://github.com/hw-sw-contracts/leakage-model-testing}. %

%% file: overview.tex
\section{Overview}
\label{sec:overview}

Here, we illustrate the key components of our approach.
We start by illustrating how leakage (\Cref{sect:overview:leakage}) and speculation (\Cref{sect:overview:speculation}) can be modeled using the \dsl{} language.
Next, we show how \dsl{}'s executable semantics can be used to derive leakage traces (\Cref{sect:overview:traces}).
We then overview our notion of side-channel security  (\Cref{sect:overview:side-channel-security}), and we conclude by illustrating how the \tool{} testing tool can be used to detect leaks in programs given an \dsl {} model  (\Cref{sect:overview:testing}).

\subsection{Modeling leaks with \dsl{}}\label{sect:overview:leakage}

In \dsl{}, leaks are modeled by specifying \emph{leakage clauses} (described in \Cref{sect:leakage-models}), which describe what an attacker might observe through side-channels during program execution.
As an example, \Cref{fig:ct} depicts the leakage clause formalizing   the constant-time model in \dsl{}.
We start by defining and naming the clause with the \hycode{defleakage} statement (line 1).
Next, we model the three usual requirements on constant-time programs as \emph{handlers} (lines 2--7).
Each handler consists of a guard, which specifies the operation being handled and the names of its arguments, and a body that computes leakage observations.

The first handler deals with memory load operations (lines 2--3); it assigns the target address to a new variable called \hycode{addr} and the target size (in bytes) to a new variable called \hycode{sz}. 
The body of the handler is the expression \hycode{\#("load" addr)}; it produces a tuple consisting of the string ``load'' and the target address passed to the handler. 
The second handler deals with memory store operations (lines 4--5) and is defined similarly.

The last handler (lines 6--7) targets control-flow instructions (represented as \hycode{jump}s in \dsl{}). 
The handler assigns the jump target address to \hycode{addr}, and the result of the jump condition to variable \hycode{n}. 
For unconditional jumps, \hycode{n} is simply assigned \hycode{True}. 
The body of the handler (line 7) returns a tuple consisting of the string ``jump'' and the final jump target, which we calculate by evaluating the if-then-else expression. 
If \hycode{n} is true, then the jump will be taken, so we leak the instruction target \hycode{addr}.
Otherwise, we leak the address of the next instruction calculated using the special \emph{context variables} \hycode{\&pc} and \hycode{\&insn}, where \hycode{\&pc} holds the current instruction's address  and \hycode{\&insn.size} contains its size.

\begin{figure}\begin{minted}[fontsize=\small, linenos,numbersep=0pt]{hylang}
    (defleakage ConstantTime [] 
      (on [(load [addr]_sz)
           #("load" addr)]
          [(store [addr]_sz := val)
           #("store" addr)]
          [(jump addr : n)
           #("jump" (if n addr (+ &pc &insn.size)))]))
      \end{minted}
      \label{sec:ct-model}
    \caption{Modeling constant-time requirements in \dsl{}}
      \label{fig:ct}
    \end{figure}

We describe other examples of leakage clauses in
    \cref{sect:leakage-models}.

\subsection{Modeling speculation with \dsl{}}\label{sect:overview:speculation}
By default, \dsl{} assumes that programs are executed following the
instruction set architectural semantics. However, processors achieve
significant performance improvements by speculating on the values of
intermediate computations and continuing the execution based on these
predictions. Speculation does not affect the correctness of the
computation, because the processor always checks the correctness
of its guesses and squashes all speculative execution steps in case of
a misprediction. However, the microarchitectural effects of such
instructions are not reversed; the resulting information leakage can
be exploited to recover secrets~\cite{spectre}. 

To model the effects of new leakage models under speculation, \dsl{} also allows modeling the effects of speculatively executed instructions.
For this, \dsl{} relies on \emph{prediction clauses} (described in
\Cref{sect:execution-models}) describing (a) which instructions might
trigger speculation and (b) how the predicted values (for control or
data flows) are computed.
As an example, \Cref{fig:v1} depicts a prediction clause capturing the
effects of speculating over branch instructions~\cite{spectre}.
Prediction clauses are specified using \hycode{defpredictor} statements
(line 1) and, similarly to leakage clauses, they are defined using
handlers.
Here, however, a handler's body models the set of predicted values
used when speculatively executing instructions.
For instance, the \hycode{branchSpec} clause consists of a single
handler (line~2--4) that deals with control-flow instructions
\hycode{jump} with target \hycode{addr} and condition \hycode{n}.
The handler uses a {\bf\texttt{when}} block to test whether the
current instruction is a conditional branch (line 3).

Importantly, the prediction clause \emph{only} specifies the predicted values; it does not specify how speculative instructions are squashed or under which conditions the processor checks for misspeculation.
These aspects are automatically handled by \dsl{}, which will speculatively explore all wrong predictions produced by the prediction clause following the  \emph{always mispredict} strategy~\cite{spectector}.
In our example, the handler computes the mispredicted target in line 4 (where \hycode{"PC"} indicates  a control-flow prediction).
\dsl{}, then, will use the target computed by the handler to
speculatively explore, for a fixed number of steps, the mispredicted
branch.\looseness=-1

In \Cref{sect:execution-models}, we describe in detail how prediction
clauses work and model in \dsl{} different forms of speculative execution.

\begin{figure}\begin{minted}[fontsize=\small, linenos,numbersep=0pt]{hylang}
    (defpredictor branchSpec []
    (on [(jump addr : n)
         (when (&insn.group CS_GRP_JUMP)
           [("PC" (if n (+ &pc &insn.size) addr))])]))
      \end{minted}
    \caption{Modeling branch speculation in \dsl{}}
      \label{fig:v1}
    \end{figure}

\subsection{Generating leakage traces}\label{sect:overview:traces}
Testing for leakage with respect to a given \dsl{} model requires
deriving \emph{leakage traces}, following the model's leakage and
prediction clauses, directly from program executions.
For this, we implemented an \emph{executable version of \dsl{}} on top
of the Unicorn CPU emulator~\cite{unicorn}, which allows simulating
architectural executions.
Given an \dsl {} model, we automatically compile leakage and execution
clauses into Unicorn \emph{event hooks}.
These hooks are used to instrument program execution whenever
user-defined architectural events are triggered.
For leakage clauses, \dsl{} handlers are translated into hooks
monitoring events and generating leakage observations.
For instance, the \hycode{load} and \hycode{store} handlers from \Cref{fig:ct} are translated into hooks intercepting memory requests and recording the address as part of the leakage trace.
For prediction clauses, \dsl{} handlers are translated into hooks that instrument program execution to explore speculative paths in addition to the architectural execution.
As an example, the clause from \Cref{fig:v1} is translated into a hook that starts the speculative execution of the mispredicted branch.
Thus, the executable version of \dsl{} allows us to directly derive the leakage traces associated with each program execution for arbitrary x86 binaries.
We provide further details on \dsl{}'s executable version in \Cref{sect:testing-approach:dsl-executable-semantics}.

\subsection{Specifying side-channel security}\label{sect:overview:side-channel-security}

As standard, we formalize side-channel security as \emph{non-interference} with respect to a leakage model~\cite{almeida2016verifying}.
Informally, a program is secure if its leakage traces
do not leak secrets. 
In more detail, the notion of security is parametrized by a 
\emph{labeled interface} for the program under consideration.
The interface declares, for each program argument, its security level (secret or public) and additional constraints such as size. 
This interface defines a notion of \emph{valid input}, and a notion of \emph{low-equivalence} between two valid inputs: informally, two inputs are low-equivalent if they only differ in their secrets. 
Then, a \emph{program is non-interfering} iff executing the program on pairs of low-equivalent valid inputs yield equal leakage traces.

\subsection{Detecting leaks with \tool{}}\label{sect:overview:testing}

Following our formalization of side-channel security, security
violations are pairs of program executions that (i) yield different
leakage traces; and (ii) start from valid low-equivalent inputs, i.e.,
program inputs that only differ in their secrets.

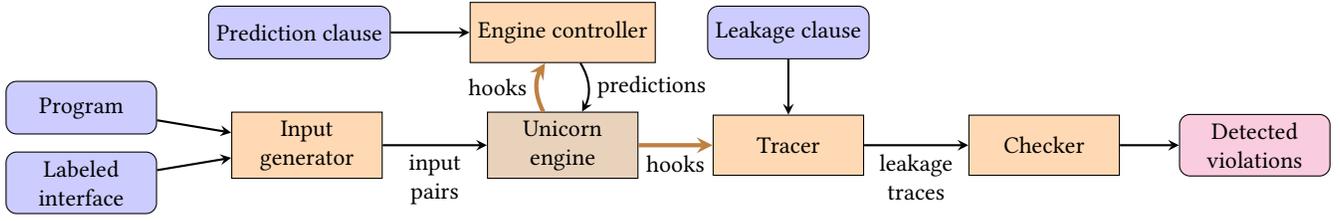
\begin{figure*}
  \input{diagram}
  \caption{Diagram showing \tool{}'s internal workflow. A user
    provides as input a leakage clause and possibly a prediction clause,
    as well as the program to test. The labeled interface specifies the format of inputs and which inputs are secret.}
  \label{fig:diagram}
\end{figure*}

To automatically discover such violations, we developed the \tool{}
testing tool, whose workflow is depicted in \Cref{fig:diagram}.
\tool{} takes as input:
\begin{inparaenum}[(a)]
\item the binary of the \textit{program} under test;
\item an entry point in a cryptographic library, with a
  \textit{labeled interface}; and
\item an \dsl {} leakage model.
\end{inparaenum}
\tool{} starts by generating random test cases, each one consisting of
a pair of valid low-equivalent inputs.
For generating a test case, \tool{} first generates a random input satisfying the constraints of the labeled interface and then re-randomizes all secret bytes to generate a public-equivalent comparison input.
Next, \tool{} derives the leakage traces associated with each input in a test case using \dsl{}'s executable semantics.
Finally, \tool{} compares the traces in a test case and reports all cases where traces differ, i.e., indicating secret-dependent leaks.

%% file: diagram.tex
\tikzstyle{input} = [rectangle, rounded corners, minimum width=2cm, minimum height=.7cm,text centered, draw=black, fill=blue!20]
\tikzstyle{output} = [rectangle, rounded corners, minimum width=2cm, minimum height=.7cm,text centered, draw=black, fill=magenta!25]
\tikzstyle{spec} = [rectangle, rounded corners, minimum width=2cm, minimum height=.7cm,text centered, draw=black, fill=blue!20]
\tikzstyle{process} = [rectangle, minimum width=2cm, minimum height=.8cm, text centered, draw=black, fill=orange!30]
\tikzstyle{external} = [rectangle, minimum width=2cm, minimum height=.8cm, text centered, draw=black, fill=brown!35]
\tikzstyle{arrow} = [thick,->,>=stealth]
\tikzstyle{hook} = [draw=brown,ultra thick,->,>=stealth]

\begin{tikzpicture}[node distance=3cm]
  \node (program) [input] {Program};
  \node (interface) [input, below of = program, yshift=2cm, align=center] {Labeled \\ interface};
  \node (inputgen) [process, right of = interface, yshift=.5cm, align=center] {Input \\ generator};
  \node (unicorn) [external, right of = inputgen, xshift=.4cm, align=center] {Unicorn \\ engine};
  \node (runner) [process, above of = unicorn, yshift=-1.5cm] {Engine controller};
  \node (exemodel) [spec, left of = runner, xshift=-.5cm] {Prediction clause};
  \node (tracer) [process, right of = unicorn, align=center] {Tracer};
  \node (leakmodel) [spec, above of = tracer, yshift=-1.5cm] {Leakage clause};
  \node (checker) [process, right of = tracer, xshift=.4cm] {Checker};
  \node (output) [output, right of = checker, align=center, xshift=-.2cm] {Detected \\ violations};

  \draw [arrow] (program) -- (inputgen);
  \draw [arrow] (interface) -- (inputgen);
  \draw [arrow] (inputgen) -- node [anchor=north, align=center] {input \\ pairs} (unicorn);
  \draw [hook] (unicorn) to [bend left] node [anchor=east] {hooks} (runner);
  \draw [arrow] (runner) to [bend left] node [anchor=west] {predictions} (unicorn);
  \draw [hook] (unicorn) -- node [anchor=north] {hooks} (tracer);
  \draw [arrow] (tracer) -- node [anchor=north, align=center] {leakage \\ traces} (checker);
  \draw [arrow] (exemodel) -- (runner);
  \draw [arrow] (leakmodel) -- (tracer);
  \draw [arrow] (checker) -- (output);

\end{tikzpicture}

%% file: formal.tex
\newcommand{\p}[1]{{\rm\it#1}}
  \newcommand{\pt}[1]{\texttt{#1}}
  \newcommand{\sym}[1]{\uwave{{\small\rm#1}}}
 \newcommand{\outp}[1]{\raisebox{-.15ex}{\large#1}}
  \newcommand{\inp}[1]{\raisebox{.7ex}{\rm\footnotesize#1}}
 \newcommand{\tn}[1]{\textnormal{#1}}
 \newcommand{\lgrp}{\inp{[}}
  \newcommand{\rgrp}{\inp{]}}
  \newcommand{\syngrp}[1]{\shat{#1}}
  \newcommand{\altdash}{\raisebox{.15ex}{\rm\footnotesize\big|}\xspace}

\section{Modeling leakage in \dsl{}}
\label{sect:leakage-models}

Here, we first introduce the \dsl{} domain specific language with a focus on leakage clauses (\Cref{sect:leakage-models:dsl}); prediction clauses are the focus of~\Cref{sect:execution-models}.
Next, we illustrate the core features of \dsl{} by modeling three microarchitectural optimization proposals from~\cite{pandorasbox}:
silent store suppression (\Cref{sect:leakage-models:silent-stores}), register file compression (\Cref{sect:leakage-models:register-compression}), and computation reuse (\Cref{sect:leakage-models:computation-reuse}).

\subsection{The \dsl{} language}
\label{sect:leakage-models:dsl}

\begin{figure}[t] \bf\raggedright
    
    \p{initializer-pair} \tn{:} \sym{name} \p{expr} \\
    \p{uop-binding} \tn{:} \\
    \hphantom{X \altdash }\pt{(read \sym{reg})}   \\
    \hphantom{X} \altdash \pt{(write \sym{reg} := \sym{val})}  \\
    \hphantom{X} \altdash \pt{(expr (\sym{op} \syngrp{ \sym{val} }\raisebox{-.3ex}{\inp{*}}))}  \\
    \hphantom{X} \altdash \pt{(addr \sym{base} + \sym{index} * \sym{scale} + \sym{off})}  \\
    \hphantom{X} \altdash \pt{(load [\sym{addr}]\uline{~}{\sym{sz}})}  \\
    \hphantom{X} \altdash \pt{(store [\sym{addr}]\uline{~}{\sym{sz}} := \sym{val})}  \\
    \hphantom{X} \altdash \pt{(jump \sym{addr} : \sym{n})}  \\
    \p{leakage-def} \tn{:} \\
    \hphantom{X}\pt{(defleakage \sym{LeakageModel} [\syngrp{ \p{initializer-pair} }\inp{*}] }\\[2pt]
    \hphantom{XX{ }}\pt{(on \syngrp{\,\pt{[\,\p{uop-binding}\;\,\p{body}$_{\;\!\!\mathit{obs}}$\,]}\,}\inp{*}))} \\
    \p{predictor-def} \tn{:} \\
    \hphantom{X}\pt{(defpredictor \sym{PredictorModel} [\syngrp{ \!\p{initializer-pair} \!}\inp{*}] }\\[2pt]
    \hphantom{XX{ }}\pt{(on \syngrp{\,\pt{[\,\p{uop-binding}\;\,\p{body}$_{\;\!\!\mathit{preds}}$\,]}\,}\inp{*}))}
    \caption{Syntax of \dsl{}. Words and symbols in {\bf\pt{fixed-width}} are verbatim keywords,
  words in \p{italics} are syntactic forms either of Hy or defined here,
  and names with \sym{underwave} can be replaced by any valid
  identifier.
      Syntactic groups with a \syngrp{ square hat } are either
      optional\inp{?} or repeatable\inp{*}.}
    \label{fig:defleakage-def}
  \end{figure}

We developed the \dsl{}  domain-specific language to specify leakage models.
\dsl{} is implemented in Hy~\cite{hy}, a Lisp-like language that compiles to Python bytecode.
This provides our models with the full expressiveness of Python as well as access to the data structures from the Python standard library, allowing users to explore different leakage models with minimal overhead.

The syntax of leakage models is given in \cref{fig:defleakage-def}. An \dsl{} leakage model consists of a leakage clause (defined by \hycode{defleakage} statements, explained below) and a prediction clause (defined by \hycode{defpredictor} statements, explained in \Cref{sec:execution-models}).

Each \textit{leakage clause} is defined by a name, a list of state
variables with their initial values, and a list of handlers.
The state variables are used to capture stateful leakage clauses, as
we show in our model of the computation reuse optimization in
\Cref{sect:leakage-models:computation-reuse}.
Their initial values can be defined using arbitrary Python expressions. 

Leakage handlers, instead, are used to generate sequences of leakage
observations based on specific events, which we call micro-operations,
happening during program execution.
\dsl{} consider seven different \emph{micro-operations}: 
$\hook{read}$, $\hook{write}$, $\hook{expr}$, $\hook{addr}$,
$\hook{load}$, $\hook{store}$, and $\hook{jump}$.
The $\hook{read}$ and $\hook{write}$ micro-operations refer to read and write operations on registers, whereas the $\hook{expr}$ micro-operation denotes computations.
There is also a dedicated  $\hook{addr}$ micro-operation for the computation of  memory addresses.
Next, $\hook{load}$ and $\hook{store}$ micro-operations refer to memory reads and writes.
Finally, $\hook{jump}$ micro-operations refer to (potentially conditional) control-flow changes, i.e., changes to the program counter.

A \emph{leakage handler} consists of a guard and a body.
The \emph{guard} of a leakage handler \p{uop-binding} is an operation with bindings for its arguments. 
In contrast, the \emph{body} of a leakage handler (\p{body}$_{\;\!\!\mathit{obs}}$) is a Python expression that evaluates to a tuple of values representing leakage observations, or to \hycode{None} if no observation is leaked. 
Note that bodies can update state variables, and can freely use special
context variables, such as those in \Cref{fig:context-variables}.

\begin{figure}
    \begin{tabular}{@{}lp{.85\linewidth}@{}}
    \toprule
    \textbf{Var.} & \textbf{Description}\\
    \midrule
    \hycode{\&insn} & information about the current instruction \\
    \hycode{\&mem} & current mapping of addresses to bytes \\
    \hycode{\&regs} & current mapping of registers to values \\
    \hycode{\&pc} &    current value of the program counter \\
    \bottomrule
\end{tabular}
\caption{\dsl{}'s context variables}\label{fig:context-variables}
\end{figure}

Next, we illustrate how \dsl{} captures concisely different kinds of leaks.
In \Cref{sect:case-studies}, we describe variants of these leakage clauses as well as further clauses  used in our case study.

\subsection{Silent store suppression}
\label{sec:global}\label{sect:leakage-models:silent-stores}

\emph{Silent store
suppression}~\cite{pandorasbox,silent-stores-1,silent-stores-2} is an
optimization that suppresses a write-back to memory when the
associated store operation does not change the value at that memory
location.
This could result in timing leaks, e.g., due to lowering the pressure on the CPU's pipeline.
Note that silent stores are considered, for instance, in the RISC-V architecture~\cite{waterman2019risc} and have been observed (on specific values) in some Intel CPUs~\cite{zerostore}. 

The leakage clause capturing these leaks emits an observation if the value about to be stored matches the value already in memory at the target address. 
This is formalized in \dsl{} as follows:
  \begin{minted}[fontsize=\small,linenos,numbersep=0pt]{hylang}
   (defleakage SilentStore []
     (on [(store [addr]_sz := val)
          (when (= val (&mem.read addr sz))
            #("ss" addr val))]))
  \end{minted}
  \label{sec:ss-model}
  The guard of the handler is a \hycode{store} expression (line 2).
  It binds the variable \hycode{addr} to the target address, \hycode{sz} to the value size, and \hycode{val} to the target value to be written.  
  The final part of the handler's body (line 4) returns a triple consisting of a logging string (here ``ss'' for ``Silent Store''), the current address, and the stored value.
  The first part of the handler's body is more interesting.  
  It is a conditional \hycode{when} block (line 3) that uses the special \hycode{\&mem} context  variable to retrieve values from memory. 
  In this case, the method \hycode{\&mem.read} is called with the target address \hycode{addr}  and size \hycode{sz} to get the current value in memory that the store operation would overwrite. 
  The \hycode{when} block compares this result to the store target value \hycode{val}, and generates an observation if these values are equal, i.e., a necessary condition for store suppression. %

\subsection{Register file compression}
\label{sec:op-write}\label{sec:ctx-regs}\label{sect:leakage-models:register-compression}

\emph{Register file compression}~\cite{register-compression-on-any-value,register-compression-on-zero} is an optimization that maps multiple logical (architectural) registers to the same physical (microarchitectural) register file entry when they hold the same value. 
This allows processors to perform more computations in parallel.
Its corresponding leakage model emits an observation whenever a register is updated with a value that is already stored in another register. 
This is captured by the following \dsl{} leakage clause:
\begin{minted}[fontsize=\small,linenos,numbersep=0pt]{hylang}
   (defleakage RegisterFileCompression []
     (on [(write reg := val)
          (when (and (in reg X86_64_GPRS)
                  (exists reg_i X86_64_GPRS
                          :where (!= reg_i reg)
                    (= val (&regs.read reg_i))))
            #("rfc" reg val))]))
\end{minted}
\label{sec:rfc-model}
The guard of the handler (line 2) is a \hycode{write} operation. 
It binds \hycode{reg} with the target register name and \hycode{val} with the
target value.
The body of the handler (lines 3--7) checks if \hycode{reg} is a general purpose
register---as opposed to segment registers or other special
registers---using the predicate \hycode{(in reg X86\_64\_GPRS)} (line 3)
and scans for another general purpose register that holds the
value \hycode{val} (lines 4--6). 
For this, it uses the special context variable \hycode{\&regs} to
query the logical registers by name,
and the built-in \hycode{exists} function to iterate over all general purpose register names (skipping the register currently being written).
If any other register already holds the target value, then the handler
exposes the target register and the shared value in the leakage observation (line 7).

\subsection{Computation reuse}
\label{sec:model-cr}\label{sec:op-expr}\label{sec:op-addr}\label{sec:ctx-state}\label{sect:leakage-models:computation-reuse}

\emph{Computation reuse}~\cite{computation-reuse} is an optimization that caches results of recent arithmetic instructions in a hardware memoization table.
Computation reuse thus avoids re-executing any instruction whose results are already present in the table.  
Its corresponding leakage model emits an observation if a computation is performed twice. 
This is captured by the following \dsl{} leakage clause: 
\begin{minted}[fontsize=\small,linenos,numbersep=0pt]{hylang}
    (defleakage ComputationReuse [memo (OrderedDict)]
      (on [(expr (op #* vals))
           (when (in op CACHING_OPS)
             (if (in vals (.get memo &pc #()))
               #("cr" op #* vs)
               (update memo &pc vals)))]))
\end{minted}
We first explain the initializer. 
To mimic a memoization table and to have it persist between calls to our handler, we use a \emph{state variable}---these are declared within the brackets following the model name in a \hycode{defleakage} definition as a sequence of names and initial values.
In line 1, we declare a state variable called \hycode{memo}; we initialize \hycode{memo} to an empty \hycode{OrderedDict}, which is a map data structure from Python's standard library that allows (re)ordering its keys, allowing us to simulate a simple ($n$-way) LRU cache of operand values, indexed by instruction address.%
\footnote{The choice of $n$ is left open, as we are not modeling any known processor design.  We expect processor designers and library developers to tailor leakage models to their own needs.}

Next, we turn to the guard (line 2). 
The form \hycode{expr} is used to model general arithmetic instructions. 
It uses Hy's list-assignment syntax to write \hycode{(expr (op \#* vals))}, where \hycode{op} is assigned the instruction mnemonic corresponding to the operation and \hycode{vals} is assigned the list of operand values.

We now turn to the handler's body (lines 3--6).  
It uses a predicate \hycode{(in op CACHING\_OPS)} to check if the current instruction is included in a set of common arithmetic instructions that we are explicitly caching (line 3).
Then, on line 4, it retrieves the cache way corresponding to the current instruction from \hycode{memo}, indexed by the current instruction address (\hycode{\&pc}).
If there is no such mapping yet, it returns  the empty collection literal \hycode{\#()}.
Then, if the operand list \hycode{vals} exists as an entry within the retrieved way, the body returns the tag \hycode{cr}, the operation and its operands as leakage observation.
Otherwise, the body calls a function \hycode{update} to update the \hycode{memo} cache with the current instruction address and operands.

\section{Modeling speculation in \dsl}
\label{sect:execution-models}
\label{sect:speculation-clauses}
\label{sec:execution-models}

In this section, we first show how speculation can be modeled in
\dsl{} using prediction clauses
(\Cref{sect:speculation-clauses:dsl}). These clauses allow users to
explore the interactions between speculative execution and leakage
clauses.
We then illustrate \dsl{}'s flexibility by giving examples of
control-flow (\Cref{sect:speculation-clauses:control-flow}) and
data-flow speculation (\Cref{sect:speculation-clauses:data-flow}).

\subsection{Prediction clauses in \dsl}\label{sect:speculation-clauses:dsl}
\dsl{} supports \emph{prediction clauses} that can be used to model speculative execution.
These clauses allow users to specify the possible predictions for every operation. 
For simplicity, we do not require users to specify  when the processor checks for misspeculation or how misspeculated instructions are squashed;  these checks are built-in into \dsl{}'s semantics. %
In particular, \dsl{} adopts the conservative \emph{always mispredict} approach for capturing the effects of speculative execution~\cite{spectector}.
Whenever the execution reaches an instruction that can result in control flow or data prediction (as indicated by the prediction clauses), \dsl{}'s semantics (1) executes the prediction clause to retrieve all predictions, (2) explores all executions associated with wrong predictions,\footnote{The predictions generated by the prediction clause might contain the correct value. To avoid treating it as a misprediction, \dsl{} always removes the correct value from the clause's result.} before (3) finally proceeding along the path of correct execution, i.e, the architectural path. 
 These design choices allow us to
drastically reduce the user effort for specifying speculative
execution in \dsl{}.

The syntax of prediction clauses is given
in~\Cref{fig:defleakage-def}.  Prediction clauses consist of a name, a
list of state variables with their initial values, and a list of
prediction handlers.
Prediction handlers are very similar to leakage handlers, and support flexible and generic speculation. This contrasts with prior tools~\cite{revizor,spectector}, which rely on hard-coded rules for identifying prediction points and possible mispredictions.
The core difference is that prediction handlers output a list of \emph{predicted values} for control-flow or data predictions.
A control-flow prediction consists of the \hycode{"PC"} keyword and the address of the next predicted instruction.
In contrast, a data flow prediction consists of a keyword (either \hycode{"MEM"} indicating that we are predicting a memory value or \hycode{"REG"} indicating that we are predicting a register value) and a prediction, which can be a triple \hycode{(address size value)} for memory predictions or a pair \hycode{(registerId value)} for register predictions.
These predictions are then used by \dsl{} when exploring speculative paths, i.e., whenever \dsl{} starts a new speculative path, it first selects one of the outstanding predictions and applies it to the program state before starting the speculative execution.

In the rest of this section, we show examples of how control-flow (\Cref{sect:speculation-clauses:control-flow}) and data speculation (\Cref{sect:speculation-clauses:data-flow}) can be implemented in \dsl{}.
In \Cref{sect:case-studies}, we describe other variants of the prediction clauses that we used in our case study.

\subsection{Control-flow speculation}\label{sect:speculation-clauses:control-flow}

We already presented a first example of control-flow speculation, i.e., misprediction of conditional branch instructions, in \Cref{sect:overview:speculation}.
Here, we present two other examples: straight-line speculation and speculation using a return stack buffer.

\compactparagraph{Straight-line speculation}
Some AMD processors can, under certain circumstances, execute the instructions following \emph{any} branch instruction~\cite{Wieczorkiewicz22}, including \emph{unconditional} branches.
This is captured by the following prediction clause:
\begin{minted}[fontsize=\small,linenos,numbersep=0pt]{hylang}
    (defpredictor StraightLineSpec []
      (on [(jump addr : n)
           [("PC" (+ &pc &insn.size))]]))
\end{minted}
The handler always adds a prediction for the instructions following the branch (line 3).
In certain cases, e.g., when a conditional branch is not taken, the following instruction is the correct next address.
To prevent treating this as a misprediction, \dsl{} always removes the correct outcome from the list of possible mispredictions returned by the handler.

\compactparagraph{Return stack buffer}
To speculate over return instructions, processors employ an auxiliary data structure called the return stack buffer (RSB), which  stores the return addresses of recent \instr{call} instructions and uses them as prediction for the actual return address.
We model RSB speculation (where the RSB is implemented using a circular buffer) with the following prediction clause:
\begin{minted}[fontsize=\small,linenos,numbersep=0pt]{hylang}
    (setv RSB_SIZE 16)
    (defpredictor RSBCircular [stack (* [0] RSB_SIZE)
                               idx 0]
      (on [(jump addr : n)
           (cond
             (&insn.group CS_GRP_CALL)
               (do (assoc stack idx
                          (+ &pc &insn.size))
                   (setv idx (%
             (&insn.group CS_GRP_RET)
               (do (setv idx (%
                   [("PC" (get stack idx))]))]))
\end{minted}
The code first defines a 16-entry buffer used for the RSB (line 1).
When handling jump operations, the code checks if these are calls or returns (lines 6 and 10).
In the case of calls, the code pushes the return address into the stack (lines 7--9).
For return instruction, instead, the code pops an entry from the stack using it as a predicted destination (lines 11--12).

\subsection{Data speculation}\label{sect:speculation-clauses:data-flow}

So far, we have shown how control-flow speculation can be formalized in \dsl{}. 
Here, we show an example of data speculation.
In particular, we model speculation over store bypasses (exploited in Spectre-STL attacks~\cite{spectrev4}), in which the processor speculates (possibly incorrectly) that an older \instr{store} does not conflict with a younger \instr{load} thereby accessing stale data.
This can be modeled with the following prediction clause:
\begin{minted}[fontsize=\small,linenos,numbersep=0pt]{hylang}
  (defpredictor StoreBypassSpec [buf (deque :maxlen SIZE)]
    (on [(load [addr]_sz)
         (lfor [waddr wsz val] buf
               :if (= [addr sz] [waddr wsz])
               ("MEM" (addr sz val)))]
        [(store [addr]_sz := val)
         (.append buf [addr sz (&mem.read addr sz)])]))
\end{minted}
The clause keeps track of recent \hycode{store} instructions and the contents they overwrite using the \hycode{buf} state variable, which is a buffer of size \hycode{SIZE} (defined in line 1).
In particular,  when executing a \hycode{store} micro-operation, the old overwritten value (extracted by the \hycode{(\&mem.read addr sz)} expression) is appended to \hycode{buf} (lines 6--7).
In contrast, when executing a \hycode{load} micro-operation, the code compares its address and size with those in the store buffer, returning the old value as a potential prediction (lines 2--5).

%% file: testing.tex
\section{Testing for leaks}\label{sect:testing-approach}
In this section, we introduce our approach for automatically testing the side-channel guarantees of programs against leaks captured by \dsl{} models.
We first describe our implementation of \dsl{}'s executable semantics (\Cref{sect:testing-approach:dsl-executable-semantics}) on top of the Unicorn CPU emulator~\cite{unicorn}, which enables us to derive leakage traces for arbitrary x86 program executions.
We then proceed to describe our testing approach (\Cref{sect:testing-approach:testing-tool}), which we implement in the \tool{} testing tool.

\subsection{Generating leakage traces for \dsl{}}\label{sect:testing-approach:dsl-executable-semantics}

To study the security implications associated with a given \dsl{} model for real-world cryptographic implementations, we need an automated way of deriving leakage traces directly from program executions.
To address this, we implemented an executable version of \dsl{} on top of the Unicorn emulator~\cite{unicorn}, which allows simulating x86 programs architecturally.
As mentioned in \Cref{sect:overview:traces}, we do so by translating   leakage and prediction clauses in \dsl{} into \emph{event hooks} in Unicorn.
Event hooks allow injecting arbitrary instrumentation code that is automatically  executed by the Unicorn emulator whenever specific events happen during program execution.
Our implementation is inspired by the Revizor tool~\cite{revizor}, which implements fixed leakage models on top of Unicorn using event hooks.\looseness=-1

\compactparagraph{Leakage clauses}
We compile each \dsl{} leakage clause with $n$ handlers into $n$ different event hooks that monitor the execution of the current instruction and record the associated leakage observations.
In particular, handlers of \hycode{load} and \hycode{store} micro-operations are compiled into memory hooks, which are executed whenever the emulator executes a memory request (as a result of an instruction).
Handlers for all other micro-operations, instead, are compiled into instruction hooks, which are executed whenever the emulator fetches a new instruction.
The body of a leakage handler, which records the leakage observation, is a Hy  expression.
We directly compile it into Python (using the Hy backend~\cite{hy}) as the body of the event hook.

\compactparagraph{Prediction clauses}
By default, Unicorn emulates instructions following the x86 architectural semantics.
To capture the effects of speculatively executed instructions, we extended Unicorn with an \emph{always mispredict} speculation model~\cite{spectector,revizor}.
First, we compile \dsl{} prediction clauses into event hooks.
These hooks, however, compute a list of predictions, rather than observations (as do the hooks associated with leakage clauses).
Following~\cite{revizor}, whenever the emulator executes an instruction that triggers a hook associated with a prediction clause, it (1) takes a checkpoint of the computation state, (2) retrieves the predicted values computed by the hook (and, if present, removes the correct value from the list of predictions) 
and (3) continues the (speculative) simulation based on one of the predictions.
When speculative execution terminates,\footnote{Following the always mispredict model~\cite{spectector}, this happens in one of the following cases: (1) the predefined speculation window is exhausted, (2) the execution encounters a speculation barrier, e.g., an \texttt{lfence} instruction, or (3) the program terminates.} the emulator restores the computation state, using the previously taken checkpoint, and either explores another speculative path (if there are other predictions) or restarts the architectural simulation. 

\compactparagraph{Context variables}
At every point during the simulation, the values of \dsl{} context variables (see \Cref{fig:context-variables}) are derived by inspecting the emulator's state.
For instance, a call to \hycode{\&mem.read} is translated into a call to the Unicorn's API for reading  memory.
Similarly, the \hycode{\&insn} structure is initialized by retrieving the current instruction from the emulator's state, disassembling it using the Capstone library~\cite{capstone}, and extracting the necessary information.

\compactparagraph{Deriving traces}
Given a program, an initial program state, and an \dsl{} model, the corresponding leakage trace is obtained by simulating the program execution using the Unicorn emulator extended with the event hooks obtained from the \dsl{} model.
In particular, the hooks associated with the prediction clause will trigger speculative execution whereas those associated with the leakage clause will track the leakage observations during execution.

\subsection{\tool{} testing tool}\label{sect:testing-approach:testing-tool}

Here, we describe our testing approach,  implemented in the \tool{} tool, for detecting leaks in a program given an \dsl{} model.

As anticipated in \Cref{sect:overview:testing}, we characterize side-channel security as a non-interference property~\cite{almeida2016verifying}. 
In particular, we say that \emph{program $P$ is secure under \dsl{} model $\LeakageModel$} if for all pairs of initial program states $s,s'$, if $s$ and $s'$ \emph{only differ} in their secrets, then the leakage traces (according to the model $\LeakageModel$) associated with $P$'s  execution on states $s$ and $s'$ must be the same.
Hence, a \emph{leak in program $P$ under model $\LeakageModel$} consists of a violation of this non-interference property, that is, a pair of low-equivalent program states that result in different leakage traces.

\begin{algorithm}[!htbp]
	\caption{\tool{} testing approach}\label{alg:testing}
	\begin{algorithmic}[1]
	\Require Program $P$, labeled interface $I$, \dsl{} model $\LeakageModel$, number of test cases $N$
	\State $\conf_1, \ldots, \conf_N \leftarrow \textit{genInitConfs}(I,N)$ \Comment{generate seed states} \label{line:init-seeds}
	\For{ $\conf \in \{ \conf_1, \ldots, \conf_N \}$ }
	\State $\conf' \leftarrow \textit{mutate}(\conf, I)$\Comment{mutate state}  \label{line:mutation}
	\State $t  \leftarrow \mathit{getTrace}(P,\conf,\LeakageModel)$\Comment{collect traces}\label{line:traces-start} 
    \State $t'  \leftarrow \mathit{getTrace}(P,\conf',\LeakageModel)$ \label{line:traces-end}
		\If{$t \neq t'$} \Comment{Trace comparison} \label{line:start-leakage-comparison}
			\State{\Return{Violation detected $\tup{\conf, \conf'}$}} \label{line:end-leakage-comparison}
		\EndIf
	\EndFor
	\State{\Return{No violation detected}}
	\end{algorithmic}
	\end{algorithm}

To detect leaks, we use a relational random testing approach inspired by  Revizor~\cite{revizor} and ct-fuzz~\cite{he2020ct}.
\tool{} approach for detecting leaks is summarized in \Cref{alg:testing}.
We start by generating $N$ randomly selected initial states following the user-provided labelled interface $I$ (line~\ref{line:init-seeds}).
For each state $\conf$,  we then generate a low-equivalent state $\conf'$ 
(line~\ref{line:mutation}).
Next, we derive the leakage traces associated with $\conf$ and $\conf'$ by simulating the program under test using the Unicorn emulator extended with the \dsl{} model $\LeakageModel$ as described in~\Cref{sect:testing-approach:dsl-executable-semantics} (lines~\ref{line:traces-start}--\ref{line:traces-end}).
Finally, we compare the generated traces (lines~\ref{line:start-leakage-comparison}--\ref{line:end-leakage-comparison}):
any difference in the traces is caused by a secret-dependent leak (since $\conf$ and $\conf'$ are low-equivalent).
If we detect a difference, we report it (line~\ref{line:end-leakage-comparison}).
Otherwise, we continue the testing and move to the next test case.

Before concluding, we provide further details on the labelled interface and our test generation strategy.

\compactparagraph{Labelled interfaces}
The labelled interface provides \tool{} with a description of the program inputs.
For each input, the interface describes (1)~whether it is stored in a register or in memory, (2)~its size in bytes, (3)~whether it is public or secret, and (4)~(optionally) its location in memory.
This interface is provided by the user and allows \tool{} to generate the test cases, which, as already mentioned, are pairs of low-equivalent initial states.

\compactparagraph{Generating  states}
To generate a random initial state, we instantiate each input with a sequence of randomly generated bytes of the appropriate length, as specified in the labelled interface.
This simple strategy (which might not work for structured data) is sufficient to generate valid initial states for all cryptographic implementations we tested in \Cref{sect:case-studies}, whose inputs consist of, e.g., secret keys or messages.

\compactparagraph{Mutating states}
To mutate a random state $\conf$, we replace its secret inputs (as indicated by the interface) with sequences of fresh   randomly generated bytes of the appropriate length.
This ensures that the mutated state $\conf'$ is low-equivalent to the original state $\conf$.\looseness=-1

%% file: eval.tex
\section{Evaluation and case study}\label{sect:case-studies}
In this section, we study the impact of microarchitectural
optimizations on the side-channel security of widely used
cryptographic algorithms.
As part of this case study, we identify three core research questions, which we address in the following sections:
\begin{description}
\item[RQ1] Does \dsl{} provide an expressive and concise framework for
  specifying leakage models? (\Cref{sect:case-studies:models})
\item[RQ2] Are real-world cryptographic libraries secure under the
  different models and can \tool{} detect leaks in them? (\Cref{sect:case-studies:testing}) %
\item[RQ3] Can the leaks be exploited? (\Cref{sect:case-studies:attacks})
\end{description}

\subsection{RQ1: Expressiveness of \dsl{}}\label{sect:case-studies:models}
We use \dsl{} to model 18 leakage clauses (\Cref{sect:case-studies:models:leakage}) and six prediction clauses (\Cref{sect:case-studies:models:execution}) that we implemented in \dsl{}, which combined result in 108 \dsl{} leakage models. Full implementations in \dsl{} of all our clauses are given in {Appendices~\ref{appendix:additional-leakage-clauses}--\ref{appendix:additional-prediction-clauses}}.

\subsubsection{Leakage clauses}\label{sect:case-studies:models:leakage}
We implemented in \dsl{} 18 different leakage clauses capturing the leaks introduced by eight different classes of microarchitectural optimizations, from the classic constant time model (capturing cache-related leaks), to complex optimizations like cache compression~\cite{safecracker} and prefetching~\cite{augury} as well as security-critical optimization proposals~\cite{pandorasbox}.
Implemented clauses include:
\begin{description}[leftmargin=1em,nosep]
\item[Constant-time (CT):]  We implemented the baseline model
  (denoted by CT) shown in \Cref{fig:ct}.

\item[Silent stores (SS):]  We implemented the baseline model
  (denoted by SS) shown in \cref{sec:ss-model}, and two variants.
  The SSI0 variant restricts observations only to silent stores
  where the value being written is all-zeroes~\cite{zerostore}. The
  SSI variant produces an observation only on silent stores involving
  memory locations that have already been initialized by the
  program---the latter uses initialization handlers, an \dsl{} feature
  not presented in the paper.

\item[Register file compression (RFC):] We implemented the
  baseline model (denoted by RFC) shown in \cref{sec:rfc-model}, and
  two variants. The RFC0 variant only compresses registers that are
  all-zero~\cite{register-compression-on-zero}. The NRFC variant only
  compresses \emph{narrow} values~\cite{narrow-register-compression}:
  registers whose values are less than 16 bits are compressed into
  the same physical register. For the latter, our model checks whether
  the value being written to a register is under 16 bits and emits an
  observation whenever another register also stores a value that is
  under 16 bits.

\item[Computation simplification (CS):] We implemented three
  models.
Two models (denoted respectively by CS and CST) capture
simplification strategies proposed
by~\citet{computation-simplification}, which simplify arithmetic and
logical operations on values like $0$ and $1$.
The third, CSN, captures the effects of simplifying
multiplication instructions for narrow operands (under 32 bits).\looseness=-1

\item[Operand packing (OP):]
Operand packing~\cite{operand-packing} compresses multiple in-flight instructions (of the same type) with narrow operands into a single ``compressed'' instruction that is forwarded to the execution units.
We model it by tracking the latest $n$ {\bf\pt{expr}} micro-operations during program execution, checking whether some of these micro-operations can be compressed (i.e., they all have operands that are less than 16 bits), and producing an observation if this is the case.

\item[Computation reuse (CR):]
Computation reuse optimizations~\cite{computation-reuse} cache recent computation results and avoid re-executing computations whenever their results are cached.
We implemented two models capturing the effects of reuse optimizations from~\cite{computation-reuse}.
The first model (denoted CR and presented in \cref{sec:model-cr}) captures leaks introduced by computation reuse over arithmetic operations.
The second model (denoted CRA) additionally captures leaks by computation reuse on address calculations in {\bf\pt{address}} micro-operations.

\item[Cacheline compression (CC):]
Cache compression optimizations aims at compressing cache lines, thereby increasing the amount of data that can be stored in caches.
We implemented models capturing the effects of two compression strategies: 
Frequence Pattern Compression (FPC)~\cite{vsccompression} and Base-Delta-Immidiate compression (BDI)~\cite{bdicompression}.
The former compresses several common data patterns whereas the latter compresses narrow ranges of values.
In a nutshell, both models monitor the execution of memory operations and produce an observation whenever the corresponding memory request might result in compression according to the given strategy.

\item[Prefetching (PF):]
Prefetchers aim at loading memory blocks into the cache hierarchy before these blocks are requested by instructions.
We implemented \dsl{} models capturing the effects of three different prefetching strategies: (1) next-line prefetching (PFNL)~\cite{prefetch1}, which prefetches the next memory block for any {\bf\pt{load}} operation, (2) 
stream prefetching~\cite{prefetch2} (PFS), which detects whether the program is accessing addresses at a regular stride and prefetches further memory blocks along the stride, and (3) data-dependent prefetcher (PFDD) based on behavior observed on the Apple M1 chip~\cite{augury}.
All these models (1) monitor the execution of  {\bf\pt{load}} micro-operations, (2) check whether further memory blocks need to be prefetched according to the corresponding strategy, and (3) append the prefetched memory addresses to the leakage trace.

\end{description}

\subsubsection{Prediction clauses}\label{sect:case-studies:models:execution}
We implemented in \dsl{} six different prediction clauses, the default sequential prediction clause (denoted by \Seq{} and  represented in \dsl{} by the absence of any \hycode{defpredictor} statement), and five additional clauses, capturing different speculation mechanisms.
In particular, our models cover \emph{all} speculation mechanisms that have been formalized in the literature~\cite{fabian202automatic}. The implemented models are:

\compactparagraph{Conditional branch speculation (\PHT)}
The prediction clause for this model, presented in \Cref{fig:v1}, captures speculating over branch instructions following the so-called ``always mispredicts'' semantics~\cite{spectector}.

\compactparagraph{Straight-line speculation (\SLS)}
We implemented an \dsl{} model, illustrated in \Cref{sect:speculation-clauses:control-flow},  that capturess the effects of straight-line speculation implemented in some AMD cores~\cite{Wieczorkiewicz22}.
The model always speculates for a fixed number of steps beyond any  {\bf\pt{jump}} micro-operation .

\compactparagraph{Store bypass speculation (\STL)}
The \dsl{} model illustrated in \Cref{sect:speculation-clauses:data-flow}  captures the effects of speculation over store bypasses~\cite{spectrev4}.
Following~\cite{fabian202automatic}, our model speculatively ignores issued  {\bf\pt{store}} micro-operations for a fixed number of steps. %

\compactparagraph{Return address speculation (\RSB)}
We implemented two clauses modeling speculation over return instructions. 
Both models employ a return stack buffer to determine the speculation target, but they differ in how they handle buffer under- and over-flows.
One model, denoted \RSBcirc{} and presented in \Cref{sect:speculation-clauses:control-flow}, uses a circular buffer that wraps around on over- or underflows.
The other, denoted as \RSBdrop, is inspired by the return speculation in~\cite{fabian202automatic}. It simply drops its oldest entry on overflow and halts (refuses to speculate further) on underflow.

\subsubsection{Assessment}\label{sect:case-studies:models:implementation}
In terms of expressiveness, we observe that our leakage models span a large class of microarchitectural leaks.
In particular, our leakage clauses cover both standard models like constant-time as well as advanced optimizations (cache compression and prefetching) and examples from all optimization proposals studied in~\cite{pandorasbox}.
Similarly, our prediction clauses cover multiple different speculation mechanisms and all speculation mechanisms used in state-of-the-art tools~\cite{fabian202automatic,revizor}.
To the best of our knowledge, this is the largest library of leakage models (108 models) to be implemented in a single language/tool.

Regarding the conciseness of our \dsl{} implementations, we observe that 
most  of the leakage and prediction clauses are implemented in a few
lines of \dsl{}.
For instance, the prediction clauses are all implemented in less than 10 lines of \dsl{} code each. 
We remark that \dsl{} allows directly leveraging Python's standard library and data structures, which greatly simplifies implementing more complex models (like computation reuse or cache compression). 
Overall, all the aforementioned leakage and prediction clauses can be implemented in about 500 lines of code, including Python code for handling data structures like the computation reuse cache and C code for the implementation of cache compression strategies.

\begin{table*}
\newcommand{\Lhit}{{$\spadesuit$}}
\newcommand{\Lmiss}{{$\varspadesuit$}}
\newcommand{\Jhit}{{$\varheartsuit$}}
\newcommand{\Jmiss}{{$\heartsuit$}}
\newcommand{\Nhit}{{$\clubsuit$}}
\newcommand{\Nmiss}{{$\varclubsuit$}}
\newcommand{\Chit}{{\color{red!70!black}$\vardiamondsuit$}}
\newcommand{\Cmiss}{{\color{green!60!black}$\boldsymbol{\diamondsuit}$}}
\newcommand{\Cto}{{\color{blue!70!black}\texttimes}\xspace}
\newcommand{\Ghit}{{\color{gray!70!black}$\vardiamondsuit$}}
\newcommand{\Gmiss}{{\color{gray!70!black}$\boldsymbol{\diamondsuit}$}}
\newcommand{\Gto}{{\color{gray!70!black}\texttimes}\xspace}
\caption{Results of our testing campaign.
  Filled-in diamonds \Chit{} denote that \tool{} detected a secret-dependent leak
  for a specific configuration (library, algorithm, leakage clause, prediction clause),
  while outlined diamonds \Cmiss{} denote that \tool{} detected no leaks during
  testing.
  Crosses \Cto indicate that \tool{}  timed out before completing the  testing.
  To aid readability, we visually combine results for (library, algorithm,
  leakage clause) triples if they are the same for all prediction clauses.
}\label{fig:results-auto}
\input{generated_results}

\end{table*}

\subsection{RQ2: Robustness of cryptographic libraries}\label{sect:case-studies:testing}

Here, we report the results of our analysis of the security of several real-world cryptographic libraries against the leakage models from~\Cref{sect:case-studies:models}, which we conducted using \tool{}.
In the following, we first present the test subjects (\Cref{sect:case-studies:testing:test-subjects}) and the overall experimental setup (\Cref{sect:case-studies:testing:experimental-setup}).
We conclude by presenting our results (\Cref{sect:case-studies:results}).

\subsubsection{Test subjects}\label{sect:case-studies:testing:test-subjects}
We focus on eight commonly used cryptographic algorithms that span a variety of use cases for cryptographic software:
AES and SHA512 are widely used primitives for encryption and hashing, respectively;
poly1305 and salsa20 are the default primitives used for authenticated encryption in the popular  cryptographic library libsodium; 
and ed25519 and x25519 are elliptic-curve primitives. 
We also analyze the HMAC and stream-XOR constructions to see any leakage
effects from higher-level cryptographic operations.

We test the implementations of these algorithms from five different cryptographic libraries:
libsodium~\cite{libsodium}, a popular cryptographic library written in C and designed for ease-of-use and safe defaults;
  cryptlib~\cite{cryptlib} and libnettle~\cite{nettle} as alternative libraries also written in C, to compare results between different implementations;
  Rust Crypto~\cite{RustCrypto} to examine how the high level safety guarantees provided by Rust~\cite{rust} may affect leakage results;
  and libjade~\cite{jade}, a library written in the Jasmin secure assembly language~\cite{jasmin} and \emph{verified} to be constant-time.

We compiled each library with their respective default settings and compilers.\footnote{We compiled libsodium v1.0.18 with gcc v11.4.0, cryptlib v3.4.6 with clang v14.0.0, libnettle v3.8 with clang v14.0.0, Rust Crypto using sha2 v1.10, salsa20 v0.10.2, poly1305 v0.8.0, and x25519-delak v2.0.0 all with cargo 1.73.0-nightly, and libjade v2023.05-1 with jasminc v2023.06.0.}
Since some libraries only implement a subset of the studied algorithms, this results in 25 different cryptographic implementations.
For each implementation, we manually created a wrapper function that (1) runs the algorithm implementation, and (2) annotates the algorithm inputs as secret or public for \tool{}.

\subsubsection{Experimental setup}\label{sect:case-studies:testing:experimental-setup}
For each of the test subjects, we use \tool{} to test their security against all the leakage models from \Cref{sect:case-studies:models}.
Specifically, for each test subject $\mathit{trg}$ and leakage model $M$, we use \tool{} to (a) generate 100 test cases, where each test case consists of a pair of low-equivalent initial states, (b) run $\mathit{trg}$ for all test cases and collect the leakage traces generated by the model $M$, and (c) analyze the leakage traces for leaks (i.e., checking if the traces for a single test case are different).
For each test subject and leakage model, we impose a total timeout of 240 minutes and a per-test-case timeout of 30 minutes, and stop the testing whenever one of the timeouts expires.
We ran our testing campaign on an Intel Xeon Gold 6132 running Ubuntu 22.04.2 LTS.

\subsubsection{Assessment}\label{sect:case-studies:results}

\Cref{fig:results-auto} summarizes the results of our testing
campaign. The overall finding is that \emph{all} the analyzed
implementations leak. This confirms that the optimization proposals
studied in~\cite{pandorasbox}, if implemented, could potentially
introduce security issues in modern cryptographic implementations.
The testing campaign also highlights several important observations:
\begin{itemize}[nosep,leftmargin=*]
\item For the majority of leakage clauses and implementations, \tool{}
  can already detect leaks under the \Seq{} prediction clause.
\item There are several cases, however, where \tool{} can only detect
  leaks under the speculative prediction clauses (e.g., computation
  reuse optimizations in libnettle, rust-crypto, and libjade).
This indicates that, similarly to what happens in Spectre
attacks~\cite{spectre}, the interplay between speculative execution
and microarchitectural optimizations can weaken the security of
implementations.

\item The use of memory-safe languages like Rust does not seem to
significantly improve the security against the new leakage models.
In most cases, the implementations from the rust-crypto library are as leaky as the corresponding C implementations of other libraries.

\item Constant-time programming \emph{does not} prevent leaks under these new leakage models.
Despite all analyzed implementations except AES-CBC\footnote{libnettle and cryptlib implement AES using non-constant-time lookup tables.} begin constant-time (and libjade being provably so), \tool{} still detects leaks in them.
As we show in~\Cref{sect:case-studies:attacks}, specific constant-time idioms, such as constant-time swaps, introduce leaks under several of the studied leakage clauses and result in leaks that allow recovering secret keys directly from leakage traces.

\end{itemize}

\subsection{RQ3: Exploitability of leaks}\label{sect:case-studies:attacks}

To illustrate the security relevance of the leaks identified by
\tool{}, we performed an in-depth analysis of our findings against the
libsodium implementation of the X25519 key exchange algorithm.
Our analysis highlights that in several cases we can recover a victim's secret key directly from the leakage traces resulting from  different leakage models.
We present three examples focusing on leakage models associated with register file compression, computation simplification, and silent store optimizations.
All examples exploit leaks caused by the constant-time swap implementation  introduced as part of the X25519 constant-time protections.

\begin{figure}
  \footnotesize
\begin{minted}{c}
fe25519_cswap(fe25519_limb f[5], fe25519_limb g[5],
              /*secret*/ bool b) {
    mask = (-(int64_t) b);

    x[0..5] = f[0..5] ^ g[0..5];

    x[0..5] &= mask;

    f[0..5] = f[0..5] ^ x[0..5];
    g[0..5] = g[0..5] ^ x[0..5];
}
\end{minted}
  \caption{X25519 constant-time swap from libsodium, condensed for brevity.
  The variables \texttt{f}, \texttt{g}, and \texttt{x} are each 5-element \texttt{uint64\_t} arrays.}
  \label{fig:cswap}
\end{figure}

\subsubsection{X25519 compare-and-swap}\label{sect:case-studies:attacks:cswap}
\Cref{fig:cswap} depicts the function showing a simplified version of the compare-and-swap algorithm used in X25519.
The two curve points to be swapped, \texttt{f} and \texttt{g}, are encoded as 5-element \texttt{uint64\_t} arrays.
The secret bit \texttt{b} controls whether the elements should be swapped.
Since the function is called for each
bit of the secret key, learning \texttt{b} at each iteration allows an attacker to recover the secret key.

The conditional swap is implemented without using conditional branches.
First, the condition bit \texttt{b} is expanded into 64-bit mask (line 3).
The two curve points \texttt{f} and \texttt{g} are then loaded from memory and xor'ed together into \texttt{x} (line 5),
which is then masked (line 7).
The result is xor'ed again with each structure (lines 9--10); if \texttt{b} was 0,
then the mask will also be 0 and thus \texttt{x} too will become 0; performing the xors will
leave each structure unchanged.
On the other hand, if \texttt{b} was 1, then the mask will be \texttt{0xfff...ff} and \texttt{x}
will remain unchanged. Since \texttt{x} is already the xor of each structure, xor'ing it
back into each structure will serve to swap the values.
Finally, the curve points \texttt{f} and \texttt{g} are written back to memory.

Under the constant-time leakage model, this implementation has no secret-dependent
leaks, as it contains no branches nor secret-dependent memory accesses.
Unfortunately, it still leaks under several of the leakage models from~\Cref{sect:case-studies:models}, as we show next.

\subsubsection{Register compression ($\ValueRegisterCompressionZero$, $\NarrowRegisterCompression$)}\label{sect:case-studies:attacks:rfc}
The value of \texttt{mask} is derived from the secret bit \texttt{b} and is then
used to mask \texttt{x} (line 7).
Whenever the mask is $\mathtt{0}$, the resulting operations will also
produce $\mathtt{0}$; in the $\RegisterCompression$ models, we will observe
this as a cluster of compressions to the zero register, which are recorded in the leakage traces.
This allows us
to determine whether the original condition bit was~0 or~1
in each loop iteration.

\subsubsection{Computation simplification ($\TrivialComputationSimplification$)}\label{sect:case-studies:attacks:cs}
After the temporary value \texttt{x} has been masked (line 7), it is xor'ed back
against \texttt{f} and \texttt{g} respectively to perform the swap (lines 9--10).
Whenever the \texttt{mask} is $\mathtt{0}$, the
\texttt{x} will also be $\mathtt{0}$.
However, the xor'ing $\mathtt{0}$ with any value leaves it unchanged;
thus in the $\TrivialComputationSimplification$ model,
we will observe any such xor operations as being simplified, which is recorded as part of the leakage trace.
Again, this allows us to recover the condition bit $\mathtt{b}$.

\subsubsection{Silent stores ($\SilentStores$)}\label{sect:case-studies:attacks:ss}
After the values of \texttt{f} and \texttt{g} are possibly-swapped, the
two points are written back to memory.
Whenever the values are not swapped, the memory writes on lines 9--10 do not modify the memory.
In the $\SilentStores$ model, these stores will be suppressed and produce observations in the leakage traces, which let us recover \texttt{b}.\looseness=-1

\subsubsection{Assessment}
Our analysis highlights that, for several leakage models, the leaks detected by \tool{} can be exploited to recover a secret key from the leakage traces.
In particular, the leaks are directly caused by the constant-time implementation of the compare-and-swap algorithm introduced to prevent side-channel leaks under the standard constant-time model.
We also remark that the exploited leaks are already present under the $\Seq$ prediction clause, i.e., they do not result from speculative instructions.

%% file: generated_results.tex
\begin{tabular}{l|c|ccc|ccc|ccc|c|cc|cc|ccc}
\hline
 & \multicolumn{1}{|c}{\textbf{CT}} & \multicolumn{3}{|c}{\textbf{SS}} & \multicolumn{3}{|c}{\textbf{RFC}} & \multicolumn{3}{|c}{\textbf{CS}} & \multicolumn{1}{|c}{\textbf{OP}} & \multicolumn{2}{|c}{\textbf{CR}} & \multicolumn{2}{|c}{\textbf{CC}} & \multicolumn{3}{|c}{\textbf{PF}} \\
\textbf{libsodium}~ & \large{$\mathrm{{\cdot}}$} & \large{$\mathrm{{\cdot}}$} & \large{$\mathrm{{\cdot}\textsc{i}}$} & \large{$\mathrm{{\cdot}\textsc{i}_0}$} & \large{$\mathrm{{\cdot}}$} & \large{$\mathrm{{\cdot}\raisebox{0em}{\small 0}}$} & \large{$\mathrm{{\cdot}\textsc{n}}$} & \large{$\mathrm{{\cdot}}$} & \large{$\mathrm{{\cdot}\textsc{t}}$} & \large{$\mathrm{{\cdot}\textsc{n}}$} & \large{$\mathrm{{\cdot}}$} & \large{$\mathrm{{\cdot}}$} & \large{$\mathrm{{\cdot}\textsc{a}}$} & \large{$\mathrm{\textsc{fpc}}$} & \large{$\mathrm{\textsc{bdi}}$} & \large{$\mathrm{{\cdot}\textsc{nl}}$} & \large{$\mathrm{{\cdot}\textsc{s}}$} & \large{$\mathrm{{\cdot}\textsc{dd}}$} \\
\hline
\end{tabular}

~\\

~\\

\begin{tabular}{|cl|}
  \hline
  \multirow{2}{*}{\scalebox{1.2}{\rlap{\raisebox{.3em}{{\small \rlap{\Ghit}\hspace{1.1ex}}{\scriptsize \rlap{\Ghit}\hspace{1.1ex}}{\scriptsize \rlap{\Ghit}\hspace{1.1ex}}}}\rlap{\raisebox{-.1em}{\hspace{.6ex}{\scriptsize \rlap{\Ghit}\hspace{1.1ex}}{\scriptsize \rlap{\Ghit}\hspace{1.1ex}}{\scriptsize \rlap{\Ghit}\hspace{1.1ex}}}}\hspace{2.6ex}}} & \multirow{2}{*}{results for prediction clauses:} 
    \raisebox{-.1em}[0em]{\footnotesize~~{\textlangle}S\textsc{eq}{\textrangle}~~~\footnotesize{}{\textlangle}P\textsc{ht}{\textrangle}~~~{\textlangle}S\textsc{ls}{\textrangle}}
     \\
        & \hphantom{results for prediction clauses:} 
    \footnotesize\raisebox{.1em}[0em]{\hspace{3ex}{\textlangle}S\textsc{tl}{\textrangle}~~{\textlangle}R\textsc{sb}$_\bot${\textrangle}~~{\textlangle}R\textsc{sb}$_\circ${\textrangle}}
     \\
  {\footnotesize\Chit{} / \footnotesize\Cmiss{} / \footnotesize\Cto}  & leak found / not found / time out\\
  \hline
  {\large\Chit{}\,\,{\textnormal / } \!\Cmiss}  & leak found / not found for all prediction clauses \\[.2em]
  \multirow{2}{*}{
  {\large\rlap{\large{\rlap{\Chit}\hspace{1.1ex}}}\hspace{.8ex}\rlap{\raisebox{-.2em}{{\scriptsize{}\rlap{\hspace{.1ex}\Cto}\hspace{1.1ex}}}}\hspace{.3ex}\,\;{\textnormal / }
      \!\large\rlap{\large{\rlap{\Cmiss}\hspace{1.1ex}}}\hspace{.8ex}\rlap{\raisebox{-.2em}{{\scriptsize{}\rlap{\hspace{.1ex}\Cto}\hspace{1.1ex}}}}\hspace{.3ex}}} &
    leaks found / not found for all prediction clauses; \\ & one or more speculative clauses timed out \\
  \hline
\end{tabular}

%% file: discussion.tex
\section{Discussion}\label{sect:discussion}

\compactparagraph{Scope of the models}
The goal of the leakage models presented in this paper is to capture the core aspects associated with the studied microarchitectural optimizations while (a)  enabling testing of real-world cryptographic implementations and (b) illustrating the expressiveness of the \dsl{} language.
As a result, our models simplify many aspects of modern CPUs, which might influence their faithfulness to any specific hardware.
Note also that some of our models are associated with optimization proposals rather than concrete implementations; different microarchitectural implementations of the same proposal, therefore, might result in different leakage profiles.

\compactparagraph{Scope of the results in \Cref{sect:case-studies}}
Our investigation highlighted that optimization proposals can potentially compromise the security of current cryptographic implementations, despite the use of constant-time programming.
Lifting the results of our testing campaign  to real-world CPUs, however, is only possible to the extent that our leakage models precisely capture the  microarchitectural information flows in these CPUs.
Even for the optimizations currently implemented by modern processors (e.g., prefetching), \tool{} results might incorrectly classify programs either as secure or insecure due to mismatches between actual and modeled leaks.

Regarding the exploitability analysis in \Cref{sect:case-studies:attacks}, an important caveat is that our analysis only refers to leakage models, given that the studied optimizations are (to the best of our knowledge) not yet implemented in modern CPUs.
Even considering a CPU implementing the target optimizations, lifting our analysis to a full-blown attack might be challenging for two reasons.
First, our models abstract away from many detailed aspects of modern CPU microarchitectures and, thus, might not faithfully capture all leaks happening on a specific CPU.
Second, our analysis assumes that every leakage observation is immediately visible to an attacker.
However, in a practical setting, the attacker will not have access to such precise observations.
Instead, they will only be able to observe (noisy) measurements, e.g., variations in program execution time.

\compactparagraph{Scope of \dsl{} and \tool{}}
We see \dsl{} and \tool{} as a way for both hardware vendors and cryptographic developers alike to easily study the security implications of microarchitectural optimization proposals.
This will enable identifying potential leaks during the development of new microarchitectural optimizations before their implementation in silicon, thereby enabling programmers to develop program-level countermeasures early on.

%% file: related.tex
\section{Related work}
\compactparagraph{Comparison with the Pandora works}
Here, we review~\cite{pandorasbox}, which is a direct inspiration for
our work, and a recent follow-up paper~\cite{pandora-fix}.

Sanchez Vicarte et al.~\cite{pandorasbox} conduct a systematic review
of the security implications of microarchitectural optimizations and
perform an in-depth analysis of seven classes of microarchitectural
optimizations. Their work provides semi-formal descriptions of the
leakage models associated to several optimizations (e.g., operand
packing and computation reuse). Their descriptions are based on the
notion of microarchitectural leakage descriptors (MLDs). However,
these descriptors are informal and often incomplete. For instance, the
MLD for computation reuse does not describe how the reuse buffer is
updated throughout execution. Our models, instead, capture the salient
aspects of specific optimization proposals and have executable
implementations. For instance, $\ComputationReuse$ and
$\ComputationReuseAddress$ capture the key aspects of computation
reuse from \citet{computation-reuse}.

A main novelty of our work is a framework for evaluating the security
implications of arbitrary microarchitectural proposals against
real-world cryptographic implentations by (a) modeling them as \dsl{}
leakage models, and (b) using \tool{} to automatically detect leaks
through random testing. Additionally, we consider the interactions
between the leakage models and speculative execution, which are not
explored systematically in~\cite{pandorasbox}. Our evaluation in
\Cref{sect:case-studies} considers all optimization classes
from~\cite{pandorasbox} and our results confirm that, if implemented,
such microarchitectural optimizations might compromise the security of
existing cryptographic implementations.\looseness=-1

Sanchez Vicarte et al.\ also explore the relevance of security leaks
for two classes of optimizations: silent stores and data-dependent
prefetching. Concretely, they present two proof-of-concept attacks
exploiting silent stores (against Bitslice AES128) and data-dependent
prefetching (against Ebpf). Their attack for silent-stores is
implemented on top of Gem5~\cite{lowepower2020gem5} and allows to
recover the secret key used by the Bitslice AES128 encryption
algorithm. The attack relies on (1) secret-dependent information being
copied to the stack, and (2) the attacker being able to call the
encryption algorithm repeatedly with attacker-controlled data (to
trigger silent stores). The leak exploited in this attack is similar
to several $\SilentStores$ leaks found by \tool{}, e.g., those
detected in libsodium's Salsa20 implementation. Our analysis of x25519
corroborates their findings that these leaks could be exploited in
idealized scenarios where the attacker is able to observe fully the
leakage described by the \dsl{} model.

In a follow up work~\cite{pandora-fix}, Flanders et al.\ develop a
program rewriting approach for hardening cryptographic implementations
against two of Pandora's leakages: silent stores and computation
simplification. Their approach shows that it is possible to protect
implementations against some of the Pandora leakages. However, it
incurs a significant performance overhead due to its generality. Our
work does not consider mitigations. However, it would be interesting
to explore in the future how our approach could be leveraged for
validating algorithm-specific mitigations.

\compactparagraph{Attacks related to the studied leakage models}
Here, we review existing attacks targeting leaks related with the
leakage models studied in this paper.

Ciphertext attacks~\cite{cipherleaks, LiWWETZ22} exploit the memory encryption in AMD SEV-SNP which employs a tweakable encryption mode where a ciphertext depends on a plaintext and a physical address.
Whenever a store operation at location $n$ happens, an attacker can infer whether the new value is different from the old value at that location by observing changes in the ciphertext at $n$ (ciphertexts are different iff plaintexts are different). 
This is exactly the same leakage model as for silent stores (\Cref{sect:leakage-models:silent-stores}).
Hence, mitigations against ciphertext attacks~\cite{cipherfix, cipherh} should be effective also against silent store leaks.\looseness=-1

Finally, the Augury attack~\cite{augury} exploits the 1-level pointer chasing prefetcher implemented in M1 processors, whereas the Safecracker attack~\cite{safecracker} exploits data compression schemes in caches to infer the content of cache lines.
The leakage clauses for prefetching (\textbf{PFDD}) and cache compression ($\CacheCompression$) used in our case study are inspired by the leaks exploited in~\cite{augury,safecracker}.

\compactparagraph{Attacks outside our leakage models}
Several recent works have exploited leakage through power consumption of the CPU, either directly~\cite{LippKOSECG21} or indirectly through the impact of the power consumption on the CPU heat and frequency~\cite{hertzbleed, WangPWGGFKS23, Taneja0XSGY23}.
As these attacks do not observe the microarchitecture directly, it is not clear whether they can be modeled in \dsl.

RAMBleed~\cite{KwongGGY20} exploits the Rowhammer effect~\cite{KimDKFLLWLM14} to leak data from memory. 
The attack leaks data at rest, and it is not affected by execution models. 
Thus it is less compatible with our framework.
Similarly, attacks that exploit data compression~\cite{SchwarzlBSMSG23, wang2024gpuzip, AlFardanP13} rely on vulnerabilities in software, which are out of scope for \dsl.

\compactparagraph{Formal models and analysis}
Many formal models capturing timing leaks at microarchitectural level
have been proposed.
Initially, researchers proposed models capturing leaks associated with
``constant-time''~\cite{almeida2016verifying,molnar2005program}, e.g.,
by instrumenting a program's semantics to produce leakage traces
exposing memory accesses and control-flow.
More recently, researchers have proposed models capturing leaks associated with speculatively executed instructions.
Some models~\cite{contracts,spectector,scamv,patrignani2021exorcising,fabian202automatic} extend program-level semantics with dedicated observations to capture microarchitectural side effects (like cache accesses) and capture the effects of speculatively executed instructions at high level by allowing the program semantics to explore mispredicted paths for a fixed number of steps~\cite{spectector}.
Other models~\cite{contracts,pitchfork, GuancialeBD20, blade,spectre-heretostay} rely on more complex models that explicitly capture components like pipeline stages, caches, and branch predictors. 

The \dsl{} language provides a way of rapidly prototyping and formalizing these leakage models.
As a proof of \dsl{}'s expressiveness, we  used it to capture a large class of leakage models.
In particular, beyond the constant-time model (\Cref{fig:ct}), we successfully implemented in \dsl{} (a) leakage clauses capturing the leaks induced by all optimization classes studied by \citet{pandorasbox}, and (b) speculative models capturing speculation over branch, store, and return instructions as well as straight-line speculation.
We remark that \dsl{}'s design took inspiration from prior work:
(1) the notions of leakage and prediction clauses is inspired by the models from~\cite{contracts},
(2) the modeling of speculation is inspired by the always mispredict speculative semantics from~\cite{spectector}, and 
(3) its executable implementation on top of the Unicorn emulator is inspired by the Revizor testing tool~\cite{revizor}.

\compactparagraph{Testing for leaks}
Here, we review relevant prior work on detecting leaks using testing-based approaches.

There are several approaches for detecting leaks in programs against specific leakage models.
For instance, \textsc{CtFuzz}~\cite{he2020ct} and \textsc{CtGrind}~\cite{ctgrind} detect leaks against the constant-time  model (see \Cref{fig:ct} for its \dsl{} encoding).
In particular, \textsc{CtFuzz} constructs a self-composition of the program under test with itself, which is then fuzzed for violations (i.e., by inspecting the traces produced by self-composed program) using the \textsc{Afl} fuzzer. 
This is different from \tool{}, which executes the program under test on individual inputs and compares pairs of traces.
In contrast, \textsc{CtGrind}~\cite{ctgrind} allows checking violations of constant-time on top of ValGrind~\cite{valgrind} using taint-tracking.
Finally, \textsc{SpecFuzz}~\cite{specfuzz} detects speculative bound  check bypasses (BCB) against an always-mispredict speculation model capturing speculation over branch instruction (i.e., the model in \Cref{fig:v1}).
Differently from \tool{}, however, all these approaches are tied to specific leakage models.\looseness=-1

Rather than relying on a leakage model, \textsc{dudect}~\cite{reparaz2017dude} detects side-channel leaks in programs by directly performing hardware measurements. %
Therefore, \textsc{dudect} can detect actual leaks against commercial processors.
Finally, Microwalk~\cite{WichelmannMES18, WichelmannSP022} employs binary instrumentation to collect leakage log from functions, detecting leakage when logs are affected by change in secret variables.

Testing has also been used to automatically discover leaks in processors rather than specific programs.
For instance, tools like Scam-V~\cite{buiras2021validation,scamv} and {Revizor}~\cite{revizor,oleksenko2023hide,hofmann2023speculation} %
 can be used to detect leaks in commercial processors against a leakage model used as a specification.
Other approaches~\cite{gras2020absynthe, moghimi2020medusa, weber2021osiris}, instead, detect leaks by analyzing hardware measurements without the help of a formal leakage model.
Finally, tools like SpecDoctor~\cite{hur2022specdoctor}, SIGFuzz~\cite{rajapaksha2023sigfuzz}, and AutoCC~\cite{orenes2023autocc} can test processor designs for leaks and they are applicable in the pre-silicon phase.
We remark, however, that all these tools differ in scope from \tool{}: \tool{}  {detect} leaks in programs, whereas these tools detect leaks in the CPU under test.\looseness=-1

Finally, Pensieve~\cite{yang2023pensieve} is a framework for evaluating
the security of early-stage microarchitectural defenses against Spectre attacks.
It allows hardware developers to (a)~specify a given countermeasure on top of an out-of-order processor model, and (b)~find counterexamples to speculative non-interference~\cite{spectector} using model checking.
Thus, Pensieve aims at finding problems in hardware countermeasures.
In contrast, our approach aims that evaluating the implications of microarchitectural proposals on the side-channel guarantees of programs (even beyond speculative execution).\looseness=-1

%% file: conclusion.tex
\section{Conclusion}

In the future,  chip vendors are expected to implement new and more aggressive microarchitectural optimizations to speed up computation. 
Given the extended development time of hardware mitigations, and the performance cost and often slow adoption rate of software countermeasures, it is critical that the security analysis of these new  optimizations happens \emph{early on} during their development, ideally before their availability in commercial processors.

To enable this early-stage security analysis,
we proposed a framework (consisting of the \dsl{} language and the \tool{} testing tool) for evaluating the side-channel guarantees of programs against (future) microarchitectural optimizations.
With our framework, we performed a large-scale study of the  implications of a large class of optimizations, recently identified as security-critical, on the security of mainstream cryptographic libraries.
Our results confirmed that these optimizations, if implemented, would compromise the security of all analyzed libraries.

\begin{acks}
We would like to thank Boris K\"opf for his feedback on earlier versions of this paper.

\newcounter{thesponsor}
\setcounter{thesponsor}{0}
This work was partially supported by
\stepcounter{thesponsor}the \grantsponsor{\arabic{thesponsor}}{Spanish Ministry of Science and Innovation}{https://www.ciencia.gob.es/} under the project \grantnum{\arabic{thesponsor}}{TED2021-132464B-I00 PRODIGY}; 
\stepcounter{thesponsor}the \grantsponsor{\arabic{thesponsor}}{Spanish Ministry of Science and Innovation}{https://www.ciencia.gob.es/} under the Ram\'on y Cajal grant \grantnum{\arabic{thesponsor}}{RYC2021-032614-I};
\stepcounter{thesponsor}the \grantsponsor{\arabic{thesponsor}}{Spanish Ministry of Science and Innovation}{https://www.ciencia.gob.es/} under the project \grantnum{\arabic{thesponsor}}{PID2022-142290OB-I00 ESPADA};
\stepcounter{thesponsor}the \grantsponsor{\arabic{thesponsor}}{Australian Research Council}{https://www.arc.gov.au/} Discovery Project \grantnum{\arabic{thesponsor}}{DP210102670};
\stepcounter{thesponsor}the \grantsponsor{\arabic{thesponsor}}{Deutsche Forschungsgemeinschaft (DFG, German Research Foundation)}{https://www.dfg.de/} under Germany's Excellence Strategy - \grantnum{\arabic{thesponsor}}{EXC 2092 CASA - 390781972};
\stepcounter{thesponsor}the \grantsponsor{\arabic{thesponsor}}{Air Force Office of Scientific Research (AFOSR)}{} under award number \grantnum{\arabic{thesponsor}}{FA9550-20-1-0425}
\stepcounter{thesponsor}the \grantsponsor{\arabic{thesponsor}}{Defense Advanced Research Projects Agency (DARPA)}{} under contract number \grantnum{\arabic{thesponsor}}{W912CG-23-C-0022}
\stepcounter{thesponsor}the \grantsponsor{\arabic{thesponsor}}{National Science Foundation (NSF)}{} under grant number \grantnum{\arabic{thesponsor}}{CNS-1954712}
and gifts from Intel, Qualcomm, and Cisco.
\end{acks}

%% file: leakagemodels.tex
\section{Additional leakage clauses}
\label{appendix:additional-leakage-clauses}

Here, we provide the \dsl{} modelling of the missing leakage clauses from \Cref{sect:case-studies:models:leakage}.

\subsection{$\mathbf{SSI}$}

\begin{minted}[fontsize=\small,linenos,breaklines,numbersep=0pt]{hylang}
    (defleakage SilentStoreInitializedOnly [initialized (set)]
    (on-start [model input]
      (.update initialized (sfor addr input.mem-initialized (- model.STACK addr))))
    (on [(store [addr]_sz := val)
         (let [addrs (range addr (+ addr sz))
               was-init (.issuperset initialized addrs)]
           (.update initialized addrs)
           (when (and was-init
                      (= val (&mem.read addr sz)))
             #("ss" addr val)))]))  
\end{minted}

\subsection{$\mathbf{SSI_0}$}

\begin{minted}[fontsize=\small,linenos,breaklines,numbersep=0pt]{hylang}
    (defleakage SilentStore0InitializedOnly [initialized (set)]
    (on-start [model input]
      (.update initialized (sfor addr input.mem-initialized (- model.STACK addr))))
    (on [(store [addr]_sz := val)
         (let [addrs (range addr (+ addr sz))
               was-init (.issuperset initialized addrs)]
           (.update initialized addrs)
           (when (and was-init
                      (= 0 val (&mem.read addr sz)))
             #("ss" addr val)))]))
\end{minted}

\subsection{$\mathbf{RFC0}$}

\begin{minted}[fontsize=\small,linenos,breaklines,numbersep=0pt]{hylang}
    (defleakage RegisterFileCompression0 []
    (on [(write reg := val)
         (when (and (in reg X86_64_GPRS)
                 (= val 0)
                 (exists reg_i X86_64_GPRS
                         :where (!= reg_i reg)
                   (= val (&regs.read reg_i))))
           #("rfc" reg val))]))
\end{minted}

\subsection{$\mathbf{RFCN}$}

\begin{minted}[fontsize=\small,linenos,breaklines,numbersep=0pt]{hylang}
    (setv NARROW_RFC_LIMIT (<< 1 16))
    (defleakage NarrowRegisterFileCompression []
      (on [(write reg := val)
           (when (and (in reg X86_64_GPRS)
                    (< val NARROW_RFC_LIMIT)
                    (exists reg_i X86_64_GPRS
                            :where (!= reg_i reg)
                      (< (&regs.read reg_i) NARROW_RFC_LIMIT)))
             #("rfc" reg))]))    
\end{minted}

\subsection{$\mathbf{CS}$}

\begin{minted}[fontsize=\small,linenos,breaklines,numbersep=0pt]{hylang}
    (setv ST-ADD #{"add" "shl" "sal" "shr" "sar"}
    ST-SUB #{"sub"}
    ST-MUL #{"mul" "imul"}
    ST-DIV #{"div" "idiv"}
    ST-AND #{"and" "or"}
    ST-XOR #{"xor"}
    )
    (defleakage SemiTrivialComputationSimplification []
    (on [(expr (op v1 v2))
     (when (is-any-of op
             ST-ADD :if (or (= v1 0) (= v2 0))
             ST_SUB :if (or (= v2 0) (= v1 v2))
             ST-MUL :if (or (= v1 0) (= v2 0)
                            (= v1 1) (= v2 1))
             ST-DIV :if (or (= v1 0) (= v2 1) (= v1 v2))
             ST-AND :if (or (= v1 0) (= v2 0)
                            (= v1 ALL1) (= v2 ALL1)
                            (= v1 v2))
             ST-XOR :if (or (= v1 0) (= v2 0))  ; XXX should this also have (= v1 v2)?
             )
                 ;; XXX rotations?
       #("cs" op v1 v2))]))
\end{minted}

\subsection{$\mathbf{CST}$}

\begin{minted}[fontsize=\small,linenos,breaklines,numbersep=0pt]{hylang}
    (setv ALL1 (- (<< 1 64) 1))
    (setv T-MUL #{"mul" "imul" "and"}
          T-OR  #{"or"}
          T-DIV #{"div" "idiv" "shl" "sal" "shr" "sar"})
    (defleakage TrivialComputationSimplification []
      (on [(expr (op v1 v2))
           (when (is-any-of op T-MUL :if (or (= v1 0) (= v2 0))
                               T-OR  :if (or (= v1 ALL1) (= v2 ALL1))
                               T-DIV :if (= v1 0))
             #("cs" op v1 v2))]))    
\end{minted}

\subsection{$\mathbf{CSN}$}

\begin{minted}[fontsize=\small,linenos,breaklines,numbersep=0pt]{hylang}
    (setv NARROW_CS_LIMIT (<< 1 32))
    (defleakage NarrowComputationSimplification []
      (on [(expr (op v1 v2))
           (when (and (in op ST-MUL)
                      (< v1 NARROW_CS_LIMIT)
                      (< v2 NARROW_CS_LIMIT))
             #("cs" op))]))
\end{minted}

\subsection{$\mathbf{OP}$}

\begin{minted}[fontsize=\small,linenos,breaklines,numbersep=0pt]{hylang}
    (setv OP_CTX_SIZE 200)
    (defleakage OperandPacking [ctx (deque)]
      (on [(expr (op v1 v2))
           (when (and (< v1 16) (< v2 16))
             (while (and ctx (>= (- &tick (. ctx [0][0])) OP_CTX_SIZE))
               (.popleft ctx))
             (for [[i [tick_i op_i]] (enumerate ctx)]
               (when (= op_i op)
                 (del (. ctx [i]))
                 (return #("op" op_i op)))
               (else
                 (.append ctx #(&tick op)))))]))
\end{minted}

\subsection{$\mathbf{CRA}$}

\begin{minted}[fontsize=\small,linenos,breaklines,numbersep=0pt]{hylang}
    (defleakage ComputationReuseWithAddresses [ctx (OrderedDict)
    ctx-addrs (OrderedDict)
    ctx-loads (OrderedDict)]
      (on [(expr (op #* vs))
      (when (in op CACHEING_OPS)
      (if (in vs (.get ctx &pc #()))
      (update ctx &pc vs)))]
      [(addr base + index * scale + off)
      (if (in #(base index scale off) (.get ctx-addrs &pc #()))
      (update ctx-addrs &pc [base index scale off]))]
      (if (in addr (.get ctx-loads &pc #()))
      (update ctx-loads &pc addr))]))
\end{minted}

\subsection{$\mathbf{CC-FPC}$}

\begin{minted}[fontsize=\small,linenos,breaklines,numbersep=0pt]{hylang}
    (defleakage FPCCacheCompression []
    (on [(load [addr]_sz)
         (let [block-addr (<< (>> addr CACHELINE_BITS) CACHELINE_BITS)
               block-data (&mem.read-bytes block-addr CACHELINE_SIZE)]
           #("cc" (fpc-size block-data)))]
        [(store [addr]_sz := val)
         (let [block-addr (<< (>> addr CACHELINE_BITS) CACHELINE_BITS)
               block-data (&mem.read-bytes block-addr CACHELINE_SIZE)
               offset (%
           (write-into block-data offset sz val)
           #("cc" (fpc-size block-data)))]))  
\end{minted}

\subsection{$\mathbf{CC-BDI}$}

\begin{minted}[fontsize=\small,linenos,breaklines,numbersep=0pt]{hylang}
    (defleakage BDICacheCompression []
    (on [(load [addr]_sz)
         (let [block-addr (<< (>> addr CACHELINE_BITS) CACHELINE_BITS)
               block-data (&mem.read-bytes block-addr CACHELINE_SIZE)]
           #("cc" (bdi-size block-data)))]
        [(store [addr]_sz := val)
         (let [block-addr (<< (>> addr CACHELINE_BITS) CACHELINE_BITS)
               block-data (&mem.read-bytes block-addr CACHELINE_SIZE)
               offset (%
           (write-into block-data offset sz val)
           #("cc" (bdi-size block-data)))]))
\end{minted}

\subsection{$\mathbf{PFNL}$}

\begin{minted}[fontsize=\small,linenos,breaklines,numbersep=0pt]{hylang}
    (setv POINTER_SIZE 8)  
    (setv CACHELINE_BITS 6)  
    (setv PAGE_BITS 12)  
    (defleakage NextLinePrefetch []
      (on [(load [addr]_sz)
           (let [cache-index (>> addr CACHELINE_BITS)]
             #("pf" (+ cache-index 1)))]))
\end{minted}

\subsection{$\mathbf{PFS}$}

\begin{minted}[fontsize=\small,linenos,breaklines,numbersep=0pt]{hylang}
    (setv PF_HITS 3)
    (defn diff [ns]
      (setv [ns1 ns2] (tee ns))
      (next ns2)
      (lfor [n m] (zip ns1 ns2) (- m n)))
    (defn direction-of? [page-hits]
      (when (< (len page-hits) page-hits.maxlen)
        (return 0))
      (setv diffs (diff page-hits))
      (cond (forall n diffs (> n 0)) 1
            (forall n diffs (< n 0)) -1
            True 0))
    (defleakage StreamPrefetch [all-page-hits (ddict #%
      (on [(load [addr]_sz)
           (let [cache-index (>> addr CACHELINE_BITS)
                 page-index  (>> addr PAGE_BITS)
                 page-hits (get all-page-hits page-index)
                 _ (when (not-in cache-index page-hits) (.append page-hits cache-index))
                 stream-dir (direction-of? page-hits)
                 next-cache-index (+ stream-dir cache-index)]
             (when (and stream-dir
                        (= (>> next-cache-index (- PAGE_BITS CACHELINE_BITS)) page-index))
               #("pf" next-cache-index)))]))
\end{minted}

\subsection{$\mathbf{PFDD}$}

\begin{minted}[fontsize=\small,linenos,breaklines,numbersep=0pt]{hylang}
    (setv M1PF_SIZE 20)
    (setv M1PF_PREFETCH 5)  ; we prefetch 5 elements
    (defleakage M1Prefetch [initialized (set)
                            accesses (deque :maxlen M1PF_SIZE)
                            marks (deque :maxlen PF_HITS)]
      (on-start [model input]
          (.update initialized (sfor addr input.mem-initialized (- model.STACK addr))))
      (on [(store [addr]_sz := val)
           (.update initialized (range addr (+ addr sz)))]
          [(load [addr]_sz)
           (let [val (&mem.read addr sz)]
    
             (setv stride 0)
             (for [[addr_i val_i] (reversed accesses)]
               (when (= val_i addr)
                 (.append marks addr_i)
                 (let [diffs (set (diff marks))]
                   (when (= (len diffs) 1)
                     (setv [stride] diffs)))
                 (break)))
    
             (.append accesses #(addr val))
    
             (when stride
               (setv last-aop (. marks [-1])
                     fetches [])
               (for [i (range M1PF_PREFETCH)]
                 (let [aop-el (+ last-aop (* i stride))]
                   (when (in aop-el initialized)
                     (.extend fetches [aop-el (&mem.read aop-el POINTER_SIZE)]))))
               #("pf" #* fetches)))]))
\end{minted}

\section{Additional prediction clauses}\label{appendix:additional-prediction-clauses}

Here, we provide the \dsl{} modelling of the missing prediction clauses from \Cref{sect:case-studies:models:execution}.

\subsection{R\textsc{sb}$_\bot$}

\begin{minted}[fontsize=\small,linenos,breaklines,numbersep=0pt]{hylang}
 (setv RSB_SIZE 16)
 (defpredictor RSBCircular [stack (* [0] RSB_SIZE)
                            idx 0]
   ;; RSB that drops oldest entry on overflow
   ;; and halts on underflow
   (on [(jump addr : n)
        (cond
          (&insn.group CS_GRP_CALL)
            (.append stack (+ &pc &insn.size))
          (&insn.group CS_GRP_RET)
            (if stack [(.pop stack)] [HALT]))]))
\end{minted}